\documentclass[
 reprint,
superscriptaddress,
nofootinbib,
 amsmath,amssymb,
 aps, 
 prl,
]{revtex4-2}

\usepackage{graphicx}
\usepackage{dcolumn}
\usepackage{bm}

\usepackage{subcaption}
\usepackage{orcidlink}

\usepackage{aas_macros} 

\usepackage{rotating}

\usepackage{amsmath, amssymb}
\usepackage{gensymb}
\usepackage{xcolor}
\usepackage{orcidlink}
\usepackage{physics}
\usepackage{multirow}

\usepackage{hyperref}
\hypersetup{colorlinks=True, urlcolor=violet, linkcolor=purple, citecolor=purple}

\newcommand{\flatt}{\texttt{flat} }
\newcommand{\likefive}{\texttt{like511} }
\newcommand{\ptsrc}{\texttt{ptsrc} }

\mathchardef\mhyphen="2D

\begin{document}

\title{Updated Constraints on the Injection Energy of Positrons Generating the Galactic 511 keV $\gamma$-ray line}

\author{Souradeep Das \orcidlink{0009-0006-3276-551X}}
\email{souradeepdas@iisc.ac.in}
\email{das.536@osu.edu}
\email{soura2302@gmail.com}
\affiliation{%
Research School of Astronomy and Astrophysics, Australian National University,\\
233 Mt Stromlo Road, Stromlo ACT 2611, Australia
}%
\affiliation{Center for High Energy Physics, Indian Institute of Science, C. V. Raman Avenue, Bengaluru 560012, India}
\affiliation{Department of Physics, Ohio State University, 191 West Woodruff Avenue, Columbus, OH 43210}
\author{Mark R.~Krumholz \orcidlink{0000-0003-3893-854X}}%
\email{Mark.Krumholz@anu.edu.au}
\affiliation{%
Research School of Astronomy and Astrophysics, Australian National University,\\
233 Mt Stromlo Road, Stromlo ACT 2611, Australia
}%
\author{Roland~M.~Crocker~\orcidlink{0000-0002-2036-2426}}%
\email{Roland.Crocker@anu.edu.au}
\affiliation{%
Research School of Astronomy and Astrophysics, Australian National University,\\
233 Mt Stromlo Road, Stromlo ACT 2611, Australia
}%

\author{Thomas Siegert \orcidlink{0000-0002-0552-3535}}
\email{thomas.siegert@uni-wuerzburg.de}
\author{Laura Eisenberger}
\email{laura.eisenberger@uni-wuerzburg.de}
\affiliation{Julius-Maximilians-Universit\"at W\"urzburg, Institut f\"ur Theoretische Physik und Astrophysik, Lehrstuhl f\"ur Astronomie, Emil-Fischer-Str. 31, D-97074 W\"urzburg, Germany
}%

\date{\today}

\begin{abstract}
    Even 50 years after the discovery of a positron annihilation line from the inner Galaxy, no class of astrophysical sources has emerged as a definitive explanation for both the emission morphology and flux. Positrons produced by dark matter annihilation or decay have been proposed, but the mass of any such candidate is constrained by continuum $\gamma$-ray emission at energies $>511$ keV. Earlier analyses have claimed that this emission requires that the positrons have kinetic energies less than a few MeV at injection, disfavoring both much of the dark matter parameter space and many potential compact astrophysical source classes such as pulsars. However, these constraints were not based on a full forward model of the absolute flux of the $\gamma$-ray line and continuum data, and did not marginalize over uncertainties about the relative angular distributions of the line and continuum. Here we describe an improved analysis that overcomes these limitations, and show that constraints on the injection energy are much weaker than previously claimed; even under conservative assumptions the data are consistent with initial energies up to $\sim 110$ MeV from INTEGRAL/SPI data alone, and up to $\sim 55$ MeV when including COMPTEL and EGRET data, subject to cross-normalization between them and INTEGRAL.
\end{abstract}
\maketitle

\section{Introduction}\label{sec:intro}

The 511 keV $\gamma$-ray line observed in the inner Galaxy remains a compelling mystery more than 50 years after its discovery \cite{Johnson1972, Leventhal1978, Purcell1997, Kinzer2001, Diehl2006, Prantzos+2011, Kierans+2020, Siegert2023a}.
Its luminosity implies the existence of a mechanism that produces positrons at a rate $\gtrsim 10^{43}$ s$^{-1}$ \cite{Aharonian1981, Siegert2016a, Siegert2023a}. While radionuclides produced in supernovae can plausibly supply this \cite{Aharonian1981, ChanLin1993}, such a scenario is difficult to reconcile with the observation that a significant fraction of the emission, perhaps as much as half, occurs in the Galactic bulge \cite{Knodlsedler2003, Knodlsedler2005, Weidenspointner2006, Prantzos2006, Bouchet2010, Siegert2016a} rather than in the disk where most of the young stars and supernovae occur.

These challenges have motivated a variety of alternative scenarios, including
positron sources tied to the old stellar populations that dominate the bulge \cite{Knodlsedler2005, Weidenspointner2006, Crocker2017, MeraEvans2022}, 
transport of positrons into the bulge \cite{Prantzos2006, Jean2009, Higdon2009, Lingenfelter2009,Panther2018b, Siegert2022a, DelaTorreLuque2024b}, 
pair production in jets of compact objects such as millisecond pulsars \cite{Wang2006} and microquasars \cite{Heinz2002, Guessoum2006, Siegert2016b},
and emission from compact binaries \cite{Weidenspointner2006, Weidenspointner2008a, Weidenspointner2008b, Bartels2018, Fuller2019},
as well as more exotic scenarios such as decaying or annihilating dark matter \cite{Boehm2004, Hooper2004, Ascasibar2006, Pospelov2007, Arkani-Hamed2009, Vincent2012, Ema2021, DelaTorreLuque2023},
metastable/excited dark matter \cite{Finkbeiner2007, Finkbeiner2009, Chen2009, Chen2010, Cline2011, Cline2012, Cappiello2023},
evaporating primordial black holes  \cite{Laha2019, DeRocco2019, Dasgupta2020, Keith2021, DelaTorreLuque2024c},
or a combination of these (e.g., \cite{Cai2021, DelaTorreLuque2024c, DelaTorreLuque2024d}).
Solutions involving a dark matter or old stellar origin are particularly attractive because they explain the bulge-dominated morphology of the signal \cite{Vincent2012}.

Measurements of continuum radiation around the line constrain these scenarios because positrons injected at relativistic energies interact with the interstellar medium (ISM) through bremsstrahlung, inverse Compton radiation, and ``in-flight'' annihilation to produce $\gamma$-ray continuum emission \cite{Aharonian1981, Beacom2006, Sizun2006, Sizun2007}. Higher initial energy-positrons yield stronger continuum radiation, so measurements of the continuum translate directly to upper limits on the initial positron energy. Previous analyses based on this idea have claimed upper bounds of $\sim 3~\mhyphen~7$ MeV on the positron initial energy \cite{Beacom2006, Sizun2006}, a limit that would exclude both many prospective astrophysical sources (e.g., highly relativistic jets) and much of the parameter space for models in which the positrons are produced by thermal relic dark matter such as Weakly Interacting Massive Particles (WIMPs)
\cite{Pospelov2008, Slatyer2016}.

However, the purported limits are sensitive both to the exact values of the $\gamma$-ray line and continuum flux -- which suffer from large systematic uncertainties due to the complex analysis required to estimate non-point source fluxes -- and to estimates of the non-radiative losses (e.g., Coulomb and ionization) that positrons undergo during deceleration from relativistic speeds. These uncertainties motivate us to revisit constraints on the maximum energy of injected positrons, combining an updated treatment of positron energy loss and radiation \cite{Krumholz2022}, relatively recent data from the INTEGRAL satellite processed using significantly improved techniques \cite{Siegert2016a, Diehl2018, Siegert2022b}, and full forward-modeling that properly marginalizes over uncertainties in both the data analysis and the astrophysical background. We allow for some combination of relativistic positron injection and a conventional, likely nucleosynthetic low-energy positron source. Our analysis shows that upper bounds on the relativistic injection energy are significantly weaker than previously claimed even when they produce a significant fraction of 511 keV emission, making positrons with injection energies up to $\sim$110 MeV fully consistent with the data.

\section{\texorpdfstring{$\gamma$-ray data}{Gamma-ray data}}

The majority of our data come from the SPI instrument aboard the INTEGRAL satellite, which offers 22 broad bins from 30 keV to 8 MeV and two narrow bins centered at 511 keV and 1809 keV to measure the line fluxes due to thermalized positron annihilation and $^{26}\rm{Al}$ decay, respectively \cite{Siegert2022b}. Because higher-energy $\gamma$-rays  help constrain higher injection energies \cite{Churazov2011}, we supplement these data with measurements from COMPTEL and EGRET aboard the CGRO satellite, which provide 3 broad bands from 10 MeV to 300 MeV \cite{Sizun2006, Strong1999}. (We use CGRO data rather than more recent data from the LAT instrument aboard the Fermi satellite because the energies of interest to us lie at the edge of LAT's sensitivity range, where its angular resolution becomes very poor -- $\approx 10^\circ$ at 20 MeV \cite{Atwood2009, Ackermann2013} -- substantially worse than CGRO's resolution of $\approx 2^\circ$ at the same energy.) It is advantageous for our analysis to use COMPTEL and EGRET data points -- using only INTEGRAL data generates the least stringent limits, up to around 100 MeV as we see in the Supplemental Material \cite{suppmat}, insufficient to distinguish between models with lower injection energies.

A central challenge when combining these data sets is that uncertainties in SPI's background level and photon arrival direction, and to a lesser extent COMPTEL's and EGRET's, make it impossible to extract the total diffuse flux within some specified angular region of the sky in a fully model-independent way.
This in turn means that we cannot be sure of the relative normalization of the SPI and CGRO data, a complication that was ignored in earlier analyses. This relative normalization is critical because, as noted above, constraints on the maximum energy of injected positrons come primarily from the relative strengths of the line and continuum.

To handle this uncertainty we use SPI fluxes measured over four regions of interest (ROIs); three circular regions centered on the Galactic center with radii of $5^\circ$, $9^\circ$, and $18^\circ$, and a square region defined by the Galactic coordinates $|l| < 47.5^\circ$ and $|b| < 47.5^\circ$. The fluxes for circular regions are obtained by fitting spatial models to the square region, which we obtain from ref. \cite{Siegert2022b}. These regions cover a range of contributions from the bulge-like and disk-like components of the 511 keV signal, and thus yield a range of line-to-continuum ratios for the SPI data. For the COMPTEL and EGRET data, we use the solid angle-integrated fluxes obtained by ref.~\cite{Sizun2006} for a  square region defined by $|l|<10^\circ$, $|b|<10^\circ$. Since CGRO/COMPTEL and CGRO/EGRET data do not cover an area that precisely matches any of the ROIs covered by the SPI analyses, we consider three possible strategies to assign these higher-energy fluxes to each ROI: \texttt{flat}, whereby we assume that the CGRO fluxes scale linearly with the solid angle of the ROI (corresponding to the fluxes being uniform on the sky), \texttt{ptsrc}, whereby we assume that the CGRO fluxes are independent of the solid angle (corresponding to the case where the fluxes arise from a point source at the Galactic center), and \texttt{like511}, whereby we assume that the ratio of the CGRO fluxes for different ROIs are identical to the ratios of the 511 keV line fluxes (corresponding to what we would expect if the high-energy fluxes were produced by exactly the same population as produces the 511 keV line, as would be the case for example in a model where both the line and the continuum are produced primarily by dark matter annihilation). Below we fit our model for each possible combination of ROI and scaling strategy. The \texttt{flat} and \texttt{ptsrc} cases represent extreme limits, while the \texttt{like511} case represents a compromise between them.

\section{Forward model for the spectrum}\label{sec:model}

\paragraph{Lepton injection model.}\label{subsec:injection} 
Our model includes two sources of leptons: mildly-relativistic positrons injected by $\beta^+$ decay of radioactive nuclei, and relativistic $e^+/e^-$ pairs from some other mechanism whose energies at injection $E_\mathrm{inj}$ we seek to fit. We neglect the fraction $\alpha$ of cases where internal bremsstrahlung (IB) yields injection energies $<E_\mathrm{inj}$, but we do include the IB emission -- see below. We divide $\beta^+$ decay into positrons produced by $^{26}$Al and by all other candidate astrophysical nuclides (e.g., $^{44}$Ti, $^{56}$Ni), since $^{26}$Al is accompanied by production of 1809 keV photons detectable by SPI. We parameterize the total positron injection rate from all sources as $\dot{n}_{e^+}^\mathrm{inj}$, the fraction injected by the relativistic source as $f_{\rm rel}$, and the fractions injected by $^{26}\rm{Al}$ and all other $\beta^+$ sources as $f_{^{26}\rm{Al}}$ and $f_{\beta}$, respectively. We leave $\dot{n}_{e^+}^\mathrm{inj}$ and the various $f$ factors as free parameters to be fit -- see End Matter for details. Thus, the differential rate of positron injection at energy $E_i$ is
\begin{equation}\label{eq:inj_spec_pos}
    \dv{\dot{n}_{\rm e^+}^{\rm inj}}{E_i} = \dot{n}_{\rm e^+}^{\rm inj}\left[f_{\rm rel}\delta(E_i-E_\mathrm{inj})+f_{\beta}\chi^\beta(E_i)+f_{\rm ^{26}Al}\chi^{\rm Al}(E_i)\right]
\end{equation}
where $\chi^{\beta}$ and $\chi^{\rm Al}$ are the positron energy distributions produced by $^{44}\rm{Ti}$ (or $^{56}\rm{Ni}$) and $^{26}\rm{Al}$ decay, respectively, which we take from refs. \cite{ChanLin1993, Prantzos+2011}. The corresponding expression for electron injection is identical, but with $f_\beta=f_{^{26}\mathrm{Al}}=0$.

\paragraph{Leptonic emission.}\label{subsec:emission}
$\gamma$-ray emission from injected leptons occurs via four channels. First, IB during pair creation produces a $\gamma$-ray spectrum \cite{Beacom2005, Boehm2006}
\begin{eqnarray}
    \dv{\dot{n}^{\rm IB}_\gamma}{E_{\gamma}} &=& \dot{n}_{e^+}^{\rm inj} f_\mathrm{rel} \frac{\alpha}{\pi}\frac{1}{E_{\gamma}}\qty[\qty[1+\qty(\frac{s'}{s})^2]\ln\qty(\frac{s'}{m_e^2})-2\frac{s'}{s}] \nonumber \\
\end{eqnarray}
where $s=4 E_\mathrm{inj}^2$ and $s'=4E_\mathrm{inj}(E_\mathrm{inj}-E_{\gamma})$. Second, $\beta^+$-decay of $^{26}$Al produces excited $^{26}$Mg nuclei that de-excite via emission of an 1809 keV photon, yielding a spectrum $d\dot{n}_\gamma^{1809}/dE_\gamma = (f_{^{26}\rm{Al}}\dot{n}_{e^+}^\mathrm{inj}/0.82) \delta(E_\gamma-1809\,\mbox{keV})$ (corresponding to a 0.82 branching ratio for $\beta^+$-decay \cite{Prantzos+2011}), where we have approximated the line profile by a $\delta$-function since our bins do not resolve the line shape. Third, leptons traversing the ISM emit cooling radiation via bremsstrahlung, inverse Compton, and, for positrons, in-flight annihilation. Since our ROIs are large and leptonic cooling times are short, all leptons injected in a given ROI likely cool completely within it \cite{Martin2012}. Thus the steady-state $\gamma$-ray production rate via ISM cooling is
\begin{equation}\label{eq:flux-cooling}
    \dv{\dot{n}^{\mathrm{cool}, e^\pm}_\gamma}{E_\gamma} = \int \dd E_i ~ \dv{\dot{n}_{\rm e^\pm}^{\rm inj}}{E_i} \dv{n_\gamma^{\mathrm{cool}, e^\pm}}{E_\gamma}\qty(E_\gamma; E_i)
\end{equation}
where $\mathrm{d}n_\gamma^{\mathrm{cool}, e^\pm}/\mathrm{d}E_\gamma$ is the number of photons per unit energy $E_\gamma$ we would expect to be produced by a single positron/electron injected with an initial energy $E_i$ as it cools to non-relativistic energies. We compute this quantity numerically using the \textsc{criptic} cosmic ray propagation code \cite{Krumholz2022} (see End Matter and Figure S1 of Supplemental Material \cite{suppmat}); this calculation also provides $f_\mathrm{IFA}(E_i)$, the fraction of positrons of initial energy $E_i$ that suffer in-flight annihilation before thermalizing, which we will require below.
    
Finally, positrons that cool to thermalize with the ISM annihilate to photons. A fraction $1-f_\mathrm{Ps} \ll 1$ (with $f_{\rm Ps}$ again a free parameter) \cite{Siegert2016a, Jean2006, Churazov2005} of these annihilate directly, yielding two 511 keV photons, while the remainder first form \textit{positronium} (Ps) \cite{Siegert2023a, Guessoum2005}, which can be para-Ps (with probability $f_\mathrm{p} = 25\%$; \cite{OrePowell1949}) or ortho-Ps (probability $f_\mathrm{o} = 75\%$). The former also decays to two 511 keV photons, while the latter undergoes three-photon decay to form a continuum below 511 keV. The resulting total $\gamma$-ray spectrum is
\begin{eqnarray}
    \lefteqn{
    \dv{\dot{n}^{\rm th}_\gamma}{E_\gamma} =
        \dot{n}^{\rm inj}_{e^+} (1 - \overline{f}_\mathrm{IFA}) 
    \times {}
    } \;
    \\ & & 
    \left[
    2\left(1-f_\mathrm{Ps} + f_\mathrm{p}f_{\rm Ps}\right)
    \phi(E_\gamma-m_e) +
    3 f_\mathrm{o} f_{\rm Ps}\dv{P^{\rm oPs}}{E_\gamma}
    \right]\nonumber,
\end{eqnarray}
where we approximate the line as a Gaussian of width 2 keV (using the line profile $\phi$), $\dd P^{\rm oPs}/\dd E_\gamma$ is the oPs continuum profile (which we take from ref.~\cite{OrePowell1949}), and $\overline{f}_\mathrm{IFA}$ is the injection-weighted mean fraction of positrons that undergo in-flight annihilation before thermalizing.

\paragraph{Backgrounds.} In addition to emission from injected leptons, our model includes three background contributions. The first is a population of unresolved point sources (UPS) such as cataclysmic variables and coronally active stars that have been shown by Chandra observations to dominate the total emission below $\approx 100$ keV, but which are dense enough on the sky that they will be unresolved at SPI's angular resolution of $\approx 3^\circ$ \cite{Krivonos2007, Revnivtsev2007a, Revnivtsev2007b, Lutovinov2020, Berteaud2022, Calore2023}. The second is inverse Compton (IC) emission by higher-energy ($\gtrsim $ GeV) cosmic ray electrons. 
The third, which becomes significant at energies $\gtrsim 10\:\rm{MeV}$, is a contribution from the low-energy tail of photons from decay of pions produced by interactions of high-energy protons in the ISM.
We describe the background model in the End Matter.

\paragraph{Fitting method.} We transform our model for the true spectrum to observed photon fluxes $F_E$ by dividing by $4\pi D_{\rm chl}^2$, where $D_{\rm chl}$ is the effective distance to the emission in channel chl, and then applying the instrument response function of ref.~\cite{Siegert2021} to convert the model fluxes to the corresponding data space bins selected for our SPI analysis. No energy response function is available for COMPTEL or EGRET, so we apply no instrumental correction for them.  Our final model has a total of 13 free parameters (see \autoref{tab:priors} of the End Matter) which we constrain using a Markov Chain Monte Carlo fit to the data for each combination of ROI and strategy for relative scaling of CGRO and INTEGRAL fluxes; we take our final result for each ROI from whichever scaling gives the highest relative probability (see End Matter). We fit our model to the absolute fluxes in our ROIs, unlike previous approaches such as refs. \cite{Beacom2005, Beacom2006} which, through their treatment of the ROIs, implicitly assumed the \flatt model for the continuum flux.

\section{Results}\label{sec:results}

\begin{figure}[ht]
    \centering
    \includegraphics[width=0.9\linewidth]{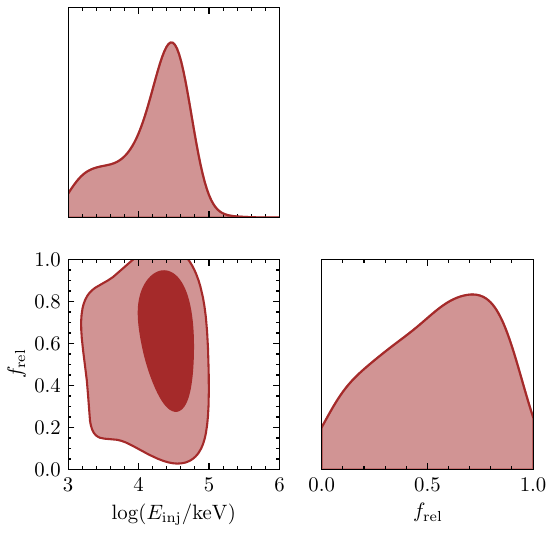}
    \caption{Posterior distribution for the pair injection energy $E_\mathrm{inj}$ (in keV) and the fraction of positrons injected relativistically ($f_{\rm rel}$) for the $9^\circ$ ROI. The shaded region in the off-diagonal subplot shows the $1\sigma$ and $2\sigma$ bounds on the joint parameter space, while the two diagonal panels show the marginal PDF for each quantity.}
    \label{fig:posterior-ptsrc-09}
\end{figure}

 \begin{figure}[ht]
    \includegraphics[width=1.0\linewidth]{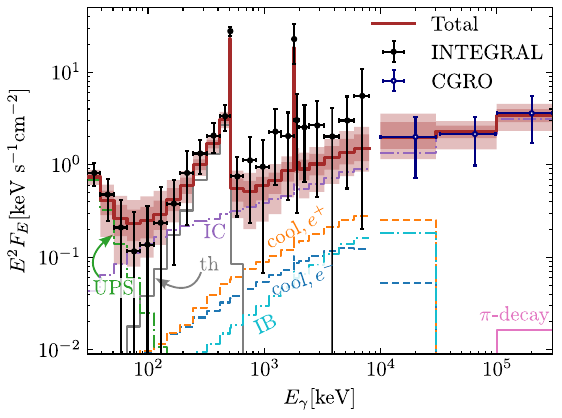}
    \caption{Fit results for our $9^\circ$ ROI. Points with error bars show observations, and the solid brown lines and shaded regions show the median and $1\sigma$ and $2\sigma$ bands of total emission produced by our MCMC fits. Other lines show median contributions from individual emission mechanisms as indicated by the labels: ${\rm cool}, e^\pm$: radiation from $e^\pm$ cooling, th: radiation from from thermalized $e^+$, IB: internal Bremsstrahlung, UPS: unresolved point sources, IC: Inverse Compton from high-energy electrons, $\pi$-decay: radiation from pion decay.}
    \label{fig:bestfit-ptsrc-09}
\end{figure}

We show a corner plot of posteriors derived from our MCMC analysis for the 9$^\circ$ ROI in \autoref{fig:posterior-ptsrc-09}; these fits are for the \texttt{ptsrc} scaling, which has marginally the highest relative probability, though for this ROI the results are similar regardless of the scaling. We provide a table of relative probabilities for all ROIs, scalings, and full corner plots for the highest-relative-probability models for each ROI in the Supplemental Material \cite{suppmat}. We show a comparison between the observed and MCMC-predicted spectra for this case in \autoref{fig:bestfit-ptsrc-09}, demonstrating that our model does a good job of reproducing the observations.

For the model shown in \autoref{fig:posterior-ptsrc-09}, corresponding to the best-fit choice \texttt{ptsrc} of cross-normalization between INTEGRAL and CGRO, we obtain a 95\% confidence upper bound  $E_\mathrm{inj} < 54$ MeV, with the relativistic source providing $f_\mathrm{rel}\approx 57\%$ of the positrons. This result is weakly dependent on the relative normalization between the two instruments. The most relaxed limit is obtained when only INTEGRAL data are used -- the upper limit relaxes further to $\sim 112~\rm{MeV}$. In principle the model also requires $E_\mathrm{inj} > 1.5\:{\rm MeV}$ and $f_\mathrm{rel} > 8\%$ at 95\% confidence, but these lower limits are driven by the need for a relativistic source to contribute to the continuum above the 511 keV line, which \autoref{fig:bestfit-ptsrc-09} shows could come equally well from electrons or positrons. Thus we cannot exclude a model with no relativistic positrons at all (i.e., the 511 keV line is entirely from $\beta^+$-decay) but with an additional source of $\sim 50$ MeV electrons.

Our results presented in the Supplemental Material \cite{suppmat} show that, for other ROIs and varying approaches to computing ISM cooling radiation, we obtain comparable upper bounds on $E_\mathrm{inj}$ and lower bounds on $f_\mathrm{rel}$. The case that yields the lowest upper bound on $E_\mathrm{inj}$ is if we assume that the positrons cool in neutral rather than ionized gas, which yields $E_\mathrm{inj} < 35$ MeV (95\% confidence). We generally find that the \texttt{ptsrc} model fits best for small ROIs and the \texttt{flat} model for larger ones, a morphology that is qualitatively consistent with reconstructions of the all-sky COMPTEL diffuse emission map \cite{Strong19a}; we discuss our findings for the best-fitting scaling model in further detail in the Supplementary Material.

\section{Discussion}\label{sec:discussion}

Our results reveal a significant region of parameter space for positron injection at relativistic energies $\sim35-55\:{\rm MeV}$ (subject to the choice of relative normalization between observations, or even up to 110 MeV in the case we disregard COMPTEL and EGRET data), which is a significant relaxation compared to previous limits of a few MeV \cite{Beacom2006, Sizun2006}. This relaxation is a result of several important advances in both data and analysis. With regard to data, previous analyses implicitly assumed that the positron annihilation line is emitted solely from the Galactic bulge, while the continuum above the line comes from both the disk and the bulge, implying that the mechanism that makes the line must produce little to no continuum. However, studies \cite{Siegert2016a} have shown that a significant fraction of the annihilation flux originates in the disk, and thus that the morphologies of the line and the continuum are much less well-separated than previously assumed. Second, unlike in previous analyses, we use a multi-component model of positron injection that includes a component from radionuclide decay, meaning only a fraction of the annihilation positrons are injected at relativistic energies. This is expected to relax the upper bounds. Finally, our calculation of the cooling radiation with \textsc{criptic} uses significantly more recent and accurate treatments of bremsstrahlung, Coulomb, and ionization losses, which cumulatively have the effect of reducing the predicted strength of cooling radiation compared to earlier work, leaving more room for higher-energy injection.

\begin{figure}
    \centering
    \includegraphics[width=1.0\linewidth]{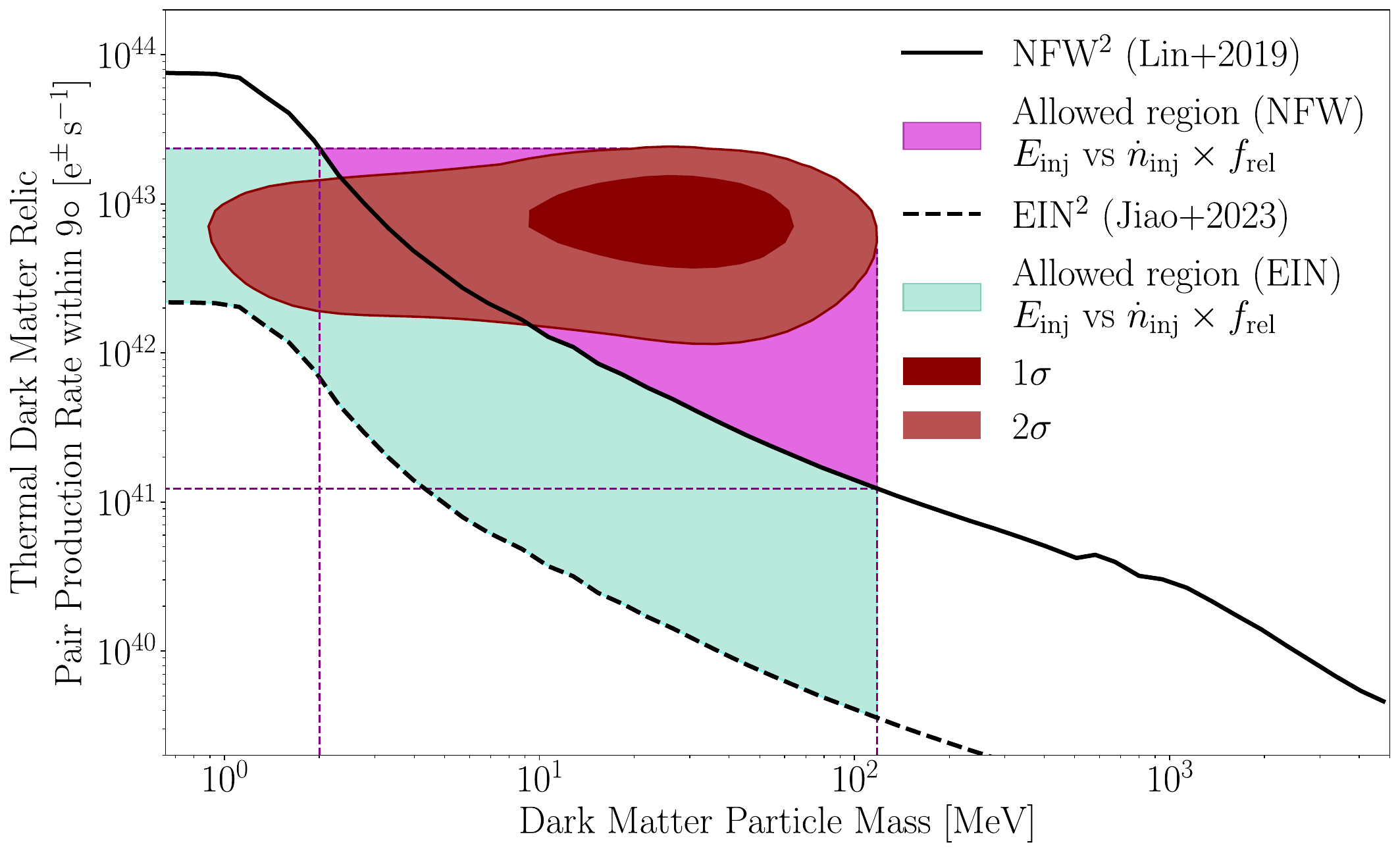}
    \caption{Thermal relic dark matter annihilation rate versus particle mass. The black solid (cusped) and dashed (cored) lines show the maximum pair production rate from thermal relic dark matter for different halo models, while contours show $1$ and $2\sigma$ confidence constraints from our fit to the 9$^\circ$ ROI. The allowed parameter space regions are shown in magenta and green, respectively.}
    \label{fig:dm_limits}
\end{figure}

To further explore the implications of our results for the thermal relic dark matter scenario \cite{Slatyer2016}, in \autoref{fig:dm_limits} we show the region of dark matter particle mass ($m_\chi=E_\mathrm{inj}$) versus annihilation rate ($\dot{n}_\mathrm{ann}=f_\mathrm{rel}\dot{n}_\mathrm{e^+}$) allowed by our fit, with the limits extended to lower values in both parameters since as we have argued we should interpret our results as providing only an upper limit on these quantities. We overplot on this the expected dark matter annihilation rate for a thermal relic with an annihilation cross-section corresponding to the 95\% C.L. upper limit from the CMB \cite{Slatyer2016} and assuming the Milky Way halo follows an NFW \cite{Lin2019} (cusped; solid line) or Einasto \cite{Jiao2023} (cored; dashed line) profile; see End Matter for computational details. For the NFW case we see that our findings allow dark matter particle masses $\approx 2$--$120$\,MeV -- the lower limit comes from the need to avoid producing too many positrons -- while for the Einasto case we only find our $\approx 120$\,MeV upper limit. However, the fact that both the Einasto and NFW lines can pass well below the best-fit region implies that dark matter with a thermal relic cross section could account only for at most 1--10\% (NFW) or 0.1--10\% (Einasto) of the $\gamma$-ray signal (or less if the annihilation branching ratio to $e^+/e^-$ pairs is less than unity), with the balance having to come from another source of relativistic leptons. 

This finding suggests pulsars as a more likely candidate for positron production, since these are expected to produce $\sim 10^{42}-10^{43}\:\rm{e^+~s^{-1}}$ in the bulge at tens of MeV \cite{Prantzos+2011}.  While it has thus long been recognized  that such magnetized compact objects could contribute significantly to the Galaxy's positron budget in principle, the relatively high characteristic energy scale, $\sim 30$ MeV \cite{deJager2009}, anticipated for the pairs escaping these objects was thought to generally disfavor them as dominant contributors. 
The loosening of the injection energy constraint we present here suggests that such sources should be re-visited, though we note that the spatial distributions of young pulsars and magnetars -- which should follow the disky distribution of Galactic star formation -- still count against these objects of sources of the bulge annihilation signal. One possible resolution to this problem would be if the sources of positrons are not primarily young pulsars but instead millisecond pulsars, which represent a much older stellar population and thus may have a significantly more bulge-like spatial distribution near the Galactic Center \citep[cf.][]{Crocker2017,Bartels2018}.

Secondary positrons produced by hadronic cosmic ray interactions with the ISM, whose characteristic injection energies also fall in the 30-40 MeV range \citep{Prantzos+2011}, represent a second candidate that had previously been discarded on energetic grounds but should now be reconsidered in light of our findings. However, like young pulsars this mechanism seems unlikely to be able to reproduce the morphology of the bulge signal, and faces the additional problem that for standard estimates of the hadronic cosmic ray population \citep{Strong2010} the expected secondary positron production is $\sim$ one order of magnitude short of what is required to explain the annihilation signal. Thus we consider millisecond pulsars a more promising model.

\begin{acknowledgments}
    The authors gratefully acknowledge useful communications with Celine Boehm, and thank Ranjan Laha for helpful comments on the manuscript. This research was undertaken with the assistance of resources from the National Computational Infrastructure (NCI Australia), an NCRIS enabled capability supported by the Australian Government, through award jh2.
    MRK and RMC acknowledge support from the Australian Research Council via Discovery Grants DP230101055 and 
    DP260100433.
    SD's visit was supported by the Future Research Talent award from the Australian National University. SD also acknowledges financial support from Kishore Vaigyanik Protsahan Yojana (KVPY) and from the Ohio State University via University Fellowship.
    LE is supported by the Bundesministerium f\"{u}r Wirtschaft und Klimaschutz via the Deutsches Zentrum f\"{u}r Luft- und Raumfahrt (DLR) under contract number 50\,OR\,2413 and is grateful for the support of the Studienstiftung des Deutschen Volkes. 
\end{acknowledgments}

\bibliography{references}

\clearpage

\section{End Matter}

\renewcommand{\theequation}{A\arabic{equation}} 
\setcounter{equation}{0} 
\renewcommand{\thetable}{A\arabic{table}} 
\setcounter{table}{0} 

\subsection{Background Emission Model}

Our model for the background to the lepton emission includes three components. The first is the population of point sources detected by Chandra that dominate emission at energies $\lesssim 100$ keV, but which are unresolved at SPI's resolution \cite{Krivonos2007, Revnivtsev2007a, Revnivtsev2007b, Lutovinov2020, Berteaud2022, Calore2023}. We model the spectral contribution from these sources as a power-law with an upper cutoff at 511 keV with functional form $d\dot{n}_\gamma^{\mathrm{UPS}}/dE_\gamma = (\mathcal{N}_\mathrm{UPS}/E_\gamma)(E_\gamma/E_0)^{\alpha_\mathrm{UPS}}$, where $E_0=50$ keV, and $\mathcal{N}_{\rm UPS}$ and $\alpha_\mathrm{UPS}$ are free parameters.

The second is inverse Compton (IC) emission from a population of electrons produced dominantly by conventional shock acceleration, which we parameterize with an energy distribution $dn_e/dE_e \propto E_e^p$ for $E_e > 1$ GeV, where the slope $p$ is a free parameter. These interact with a background radiation field consisting of three components with blackbody spectral shapes: the cosmic microwave background ($T = 2.73$ K), a far-infrared (FIR) dust radiation ($T = 30$ K), and near-infrared (NIR) starlight ($T=3000$ K). We compute the spectral shape of the resulting IC emission using the \texttt{naima} package \cite{Zabalza2015, Khangulyan+2014}, and we normalize this emission by setting the IC luminosity per unit energy at photon energy $E_\gamma = 511$ keV to a value $\mathcal{N}_\mathrm{CMB}$; this value implicitly fixes the normalization of the electron spectrum. We then normalize the IC contribution from the NIR and FIR components by defining $x_\mathrm{NIR}$ and $x_\mathrm{FIR}$ as the ratio of the energy densities in these two fields to those in the Solar neighborhood, $u_{\rm NIR} = 1$ eV cm$^{-3}$ and $u_\mathrm{FIR} = 0.5$ eV cm$^{-3}$, respectively. We leave $\mathcal{N}_\mathrm{CMB}$, $x_\mathrm{NIR}$, and $x_\mathrm{FIR}$ as parameters to be fit, which together fully specify the IC spectrum $d\dot{n}_\gamma^\mathrm{IC}/dE_\gamma$. Note that, because we allow $x_\mathrm{FIR}$ and $x_\mathrm{NIR}$ to vary, we are implicitly allowing a wide variety to spectral shapes for the radiation field with which the high-energy CRs interact to produce the background IC signal.

Our third background component is a $\gamma$-ray contribution from the decay of pions produced by nuclear inelastic collisions of protons with the ISM. We assume that the protons have a power-law momentum distribution $\dd n_p/\dd p_p \propto p_p^{q}$, and calculate the nuclear inelastic $\gamma$-ray emission as a function of $q$ directly using \textsc{criptic}, which in turn takes its treatment of this emission channel from ref.~\cite{Kafexhiu14a, Kafexhiu16a}. We leave the spectral index $q$ and the normalization $\mathcal{N}_\pi$ of the resultant photon spectrum $\dd \dot{n}_{\gamma}^{\pi}/\dd E_\gamma$ at $E_\gamma=100\:\rm{MeV}$ as free parameters to be fit.

\subsection{ISM Cooling Radiation}
 
We calculate the $\gamma$-ray emission rates $\dd \dot{n}^{\rm cool, e^\pm}_\gamma/\dd E_\gamma$ per injected cosmic ray from ISM cooling numerically from a series of simulations using the cosmic ray propagation code \textsc{criptic} \cite{Krumholz2022}, following the approach in ref.~\cite{Krumholz2023}; \textsc{criptic}'s advantages include that it provides a fully probabilistic treatment of bremsstrahlung and inverse Compton that allows realistic catastrophic losses rather than using the continuous slowing-down approximation (CSDA) and that it uses modern estimates for reaction cross sections and radiative emission. In particular, its treatment of ionization losses uses the relativistic BEQ formalism of ref.~\cite{Kim00b}, Coulomb losses uses the formalism of ref.~\cite{Gould72a}, and the treatment of bremsstrahlung (including nuclear shielding) is taken from ref.~\cite{Blumenthal70a}. In contrast, some earlier work relied on approximate analytic fitting formulae and the CSDA for these processes.

In each simulation we inject an initially mono-energetic population of electrons/positrons into a medium in which the helium abundance per H nucleon is 0.0955, and the hydrogen is assumed to be either 99\% atomic and 1\% ionized or fully ionized (our fiducial choice). The medium has both a magnetic field and a radiation field that is described as a sum of three (dilute) blackbody components with temperatures of 2.73 K, 30 K, and 3000 K (corresponding roughly to the cosmic microwave background, the dust radiation field, and the starlight field), whose energy densities are in a ratio $1 : 1.19 : 2.34$, roughly the ratio seen locally. Ref.~\cite{Krumholz2023} point out that the photon emission per injected particle $\mathrm{d}\dot{n}^{\mathrm{cool},e^\pm}/\mathrm{d}E_\gamma$ does not depend on the absolute gas density or on the energy densities of the magnetic and radiation fields. Instead, it depends only on the chemical state of the gas, the shape of the background radiation spectrum, and on two dimensionless ratios $f_\mathrm{IC}$ and $f_\mathrm{sync}$ (their Equation 11), which specify the strength of inverse Compton and synchrotron losses relative to collisional losses. For our fiducial run we adopt $f_{\rm IC} \approx f_{\rm synch}\approx 10^{-6}$. A corollary of this point is that our choice of the relative ratios of the three radiation fields that contribute to inverse Compton losses does not affect the final result, since the results are only sensitive to the sum of the energy densities of the three components, not to the energy densities of each one individually.

We run simulations at 241 initial kinetic energies $T_i$ each for both electrons and positrons; the first 201 sample points are uniformly distributed in logarithm from 100 keV to 10 MeV, and the remaining 40 points extend the grid up to 1 GeV, again with uniform logarithmic spacing from 10 MeV to 1 GeV. In each simulation particles with energy $T_i$ are injected at a steady rate, and we follow them until their kinetic energies drop below 1 keV, at which point their energies are low enough that we can treat them as thermalized. Following ref.~\cite{Krumholz2023}, we use a packet injection rate $\Gamma = 2\times 10^{-9} (n_\mathrm{H}/\mathrm{cm}^{-3})$ s$^{-1}$, where $n_\mathrm{H}$ is the background number density of H nuclei, a step size control parameter $c_\mathrm{step} = 0.05$, and a secondary production factor $f_\mathrm{sec} = 0.1$. We initially run the simulations until the time reaches $5\times (10^{9},10^{10})/(n_\mathrm{H}/\mathrm{cm}^{-3})$ s for $(T_i < 100, T_i\geq 100)$ MeV, at least five times the time required for the system to settle into steady state between particle injection and cooling. 

After this point we continue the simulations to a time $3\times (10^{10},10^{11})/(n_\mathrm{H}/\mathrm{cm}^{-3})$ s, sampling the emission at 126 uniformly-spaced times, and take $\mathrm{d}\dot{n}^\mathrm{cool,e^\pm}/\mathrm{d}E_\gamma$ for that value of $T_i$ to be the average of the instantaneous emission rates predicted at these times. We similarly measure $f_\mathrm{IFA}$, the fraction of positrons that undergo in-flight annihilation, from the difference between the positron injection rate and the rate at which positrons cool to $T < 1$ keV and we stop following them. This process therefore yields measurements of $\mathrm{d}\dot{n}^\mathrm{cool,e^\pm}/\mathrm{d}E_\gamma$ and $f_\mathrm{IFA}$ at our 241 sample values of $T_i$. For the purposes of calculating these quantities in the rest of our analysis, we simply interpolate on these tables.

Our fiducial choice of ionized composition and $f_\mathrm{sync} \approx f_\mathrm{IC} \approx 10^{-6}$ are reasonable and realistic parameters for the ISM in which positrons cool, but other choices are certainly possible. To evaluate the impact of these choices, we carry out a systematic study in which we recalculate the cooling rate using the same method but for alternative choices, and then repeat our full analysis pipeline using these alternative choices. We summarize the full set of parameters we have tested, and provide results derived from them, in Table S2 of the Supplemental Material \cite{suppmat}. As discussed in the main text, the results change little as we vary these parameters within reasonable limits, with the largest effects resulting from changing between atomic and ionized composition because Coulomb losses are slightly more efficient than ionization losses and thus an atomic medium directs somewhat more of the cooling into detectable radiation rather than invisible collisions.

\subsection{Fitting Method}\label{sec:mcmc}

\begin{table}
    \centering
    \begin{tabular}{lll}\hline\hline
       Parameter & Unit & Prior \\ \hline
    $\log \dot{n}^{\rm inj}_{e^+}/4\pi D_{\rm inj}^2$ & cm$^{-2}$ s$^{-1}$ & $\mathcal{U}(-5.0, -2.0)$ \\ 
    $\log E_\mathrm{inj}$ & keV & $\mathcal{U}(3.0, 5.48)$ \\ 
    $f_{\rm rel}$, $f_\beta$, $f_{\rm ^{26}Al}$ & - & $\rm{Dir}(1,1,1)$ \\
    $f_{\rm Ps}$ & - & $\mathcal{U}(0.9, 1.0)$ \\     
    $\log \mathcal{N}_{\rm CMB}/4\pi D_{\rm CMB}^2$ & cm$^{-2}$ s$^{-1}$ & $\mathcal{U}(-10.0, 0.0)$ \\
    $\log x_{\rm FIR}$ & - & $\mathcal{U}(-3.0, 5.0)$ \\
    $\log x_{\rm NIR}$ & - & $\mathcal{U}(-3.0, 5.0)$ \\
    $p$ & - & $\mathcal{U}(-3.5, -1.5)$ \\
    $\log \mathcal{N}_{\rm UPS}/4\pi D_{\rm UPS}^2$ & cm$^{-2}$ s$^{-1}$ & $\mathcal{U}(-4.0, 0.0)$ \\
    $\theta_{\rm UPS} = \tan^{-1}\alpha_\mathrm{UPS}$ & - & $\mathcal{U}(-\pi/2, 0)$ \\
    $\log{\mathcal{N}_{\pi} / 4\pi D^2_{\pi}}$ & cm$^{-2}$ s$^{-1}$ & $\mathcal{U}(-10.0, 5.0)$ \\
    $q$ & - & $\mathcal{U}(-3.5, -1.5)$ \\
\hline
\hline
    \end{tabular}
    \caption{Model parameters, units, and priors. Here $\mathcal{U}(a,b)$ is the uniform distribution from $a$ to $b$ and $\mathrm{Dir}(1,1,1)$ is the flat Dirichlet distribution; note that, because $f_\mathrm{rel}$, $f_\beta$, and $f_{^{26}\mathrm{Al}}$ are drawn from this distribution, and thus their sum is constrained to unity, only two of these parameters are free.}
    \label{tab:priors}
\end{table}

Our model has 13 free parameters. The leptonic emission is parametrized by: the total injection rate of positrons $\dot{n}^{\rm inj}_{e^+}/4\pi D_{\rm inj}^2$ normalized by the effective source distance $D_{\rm inj}$,
the injection energy $E_\mathrm{inj}$, 
the fraction $f_{\rm rel}$ of positrons injected by relativistic sources, the fraction $f_\beta$ injected by $\beta^+$-decay of nuclei other than $^{26}$Al (from which one derives the $^{26}$Al fraction $f_{^{26}\mathrm{Al}}=1-f_{\rm rel}-f_\beta$), and
the positronium fraction $f_{\rm Ps}$ (known to be roughly 90 -- 100\% based on models of the shape of the annihilation spectrum \cite{Siegert2016a, Jean2006, Churazov2005}).
Parameters for the background emission are: the normalization $\mathcal{N}_{\rm CMB}/4\pi D_{\rm CMB}^2$ of IC background from up-scattered CMB, the ratios $x_\mathrm{FIR}$ and $x_\mathrm{NIR}$, respectively, of the FIR and NIR radiation energy densities to that in the solar neighborhood,
the slope $p$ (known to be $\approx -3$ to $-2$, from Fermi-LAT $\gamma$-ray observations \cite{Strong2010, Siegert2022b}) of the cosmic ray electron energy distribution at energies $\geq 1$ GeV,
the normalization $\mathcal{N}_{\rm UPS}/4\pi D_{\rm UPS}^2$ and slope $\alpha_{\rm UPS}$ of the UPS background component, and the normalization $\mathcal{N}_\pi/4\pi D_\pi^2$ of the pion background component and the spectral slope $q$ characterizing the momentum distribution of the protons responsible for producing it. Note that, for fit parameters that  take the form of a luminosity or similar quantity normalized by $4\pi D^2$, we fit only a single distance-normalized quantity, not for the luminosity and distance separately. Thus for example we fit only the combination $\mathcal{N}_{\rm UPS}/4\pi D_{\rm UPS}^2$, and not $\mathcal{N}_\mathrm{UPS}$ and $D_\mathrm{UPS}$ separately.

In order to estimate constraints on the parameters, we use a Bayesian parameter inference method using the Markov-Chain Monte Carlo (MCMC) sampler \texttt{emcee} \cite{Foreman-Mackey13a}. For the purposes of this calculation we adopt the priors listed in \autoref{tab:priors}. These are mostly flat (in logarithm) over a very large parameter range. The only exceptions are: (1) for the fraction $f_\mathrm{rel}$ and $f_{\beta}$, and the implicit corresponding parameter $f_{^{26}\mathrm{Al}}$, we adopt the 3d-flat Dirichlet distribution, appropriate for three parameters constrained to have a fixed sum; (2) for $f_\mathrm{Ps}$, $p$, and $q$, we use tight priors that are uniform from 0.9 to 1 for $f_\mathrm{Ps}$ and $-3.5$ to $-1.5$ for $p$ and $q$, since these quantities are constrained by other observations -- the shape of the continuum and 511 keV line for $f_\mathrm{Ps}$ \cite{Siegert2016a, Jean2006, Churazov2005} and the slope of the cosmic ray electron and proton spectra at higher energies inferred from Fermi/LAT measurements of inverse Compton and pion decay emission \cite{Siegert2016b, Strong2010}.; (3) for $\alpha_\mathrm{UPS}$, we use the standard Jeffreys prior for slopes, whereby the prior is uniform in the angle $\theta_\mathrm{UPS} = \tan^{-1} \alpha_\mathrm{UPS}$.

We run 104 MCMC chains for 60,000 burn-in steps followed by 200,000 production steps per chain, with an estimated autocorrelation time $\approx 5000$.
To identify which scaling model best describes each ROI, we compute the Akaike Information Criterion (AIC) for each of the fits (that is, for every scaling model in each of the ROIs):
\begin{equation}\label{eq:AIC}
    {\rm AIC} = 2k - 2\log{\mathcal{L}_{\rm max}},
\end{equation}
where $k=13$ is the number of degrees of freedom of the model and $\mathcal{L}_{\rm max}$ is the maximum of the likelihood function as found by our MCMC. We compare the scaling models in each ROI using the `relative probability':
\begin{equation}\label{eq:bayes}
    p_{i} = \frac{\exp(-{\rm  AIC}_i/2)}{\sum_j \exp(-{\rm AIC}_j/2)},
\end{equation}
where $i$ and $j$ run over the three scaling models for each ROI. We accept the model with the highest relative probability in each ROI (for example, \ptsrc in the 9\degree \ region) as the most plausible scaling to adopt when combining the INTEGRAL and CGRO data. Examining the chosen fits, these tend to be the scalings that yield the smoothest transition between the lower-energy INTEGRAL data and the higher-energy CGRO bands, while scaling choices that lead to a noticeable jump between the two are disfavored.

\subsection{Thermal Relic Dark Matter Pair Production}\label{sec:dark_matter_pairs}
We calculate the electron-positron pair production rate from thermal relic dark matter as
\begin{equation}
    \Gamma_{\rm ee} = \frac{1}{2}L(\Delta\Omega) \left\langle \sigma(m_\chi) v(m_\chi)\right\rangle m_\chi^{-2}\mathrm{,}
    \label{eq:dm_pairs}
\end{equation}
where $m_\chi$ is the dark matter particle mass, $\langle \sigma(m_\chi) v(m_\chi)\rangle$ is the velocity-averaged thermal relic cross section \cite{Slatyer2016}, and $L(\Delta\Omega)$ is the luminosity of a dark matter halo that is self-annihilating, given by
\begin{equation}
    L(\Delta\Omega) = \iint_{\Delta \Omega}\,d\Omega \,\int_0^{+\infty}\,ds\,s^2\,\rho^2(s,\ell,b)\mathrm{.}
    \label{eq:los_lumi}
\end{equation}
This formalism explicitly assumes that 100\% of the dark matter annihilations go into pairs.
If other channels are allowed, such as to $\nu\bar{\nu}$ or $\gamma\gamma$, which is always the case for $m_\chi \lesssim m_\mu$, the pair production rate could in fact be even smaller.
We do not include a boost factor because boosts only become important in the outskirts of a galaxy beyond $\sim 20$\,kpc \cite{Kamionkowski2010}, well beyond our observed region.


\renewcommand{\theequation}{S\arabic{equation}} 
\setcounter{equation}{0} 
\renewcommand{\thetable}{S\arabic{table}} 
\setcounter{table}{0} 
\renewcommand{\thefigure}{S\arabic{figure}} 
\setcounter{figure}{0} 

\clearpage
\onecolumngrid

 \vspace{1em}
 
\begin{center}
  \large \textbf{Updated Constraints on the Injection Energy of Positrons Generating the Galactic 511 keV $\gamma$-ray line}\\[0.5em]
  \large \textit{\textbf{Supplemental Material}}\\[0.75em]
  \normalsize
  Souradeep Das, Mark R.~Krumholz, Roland M.~Crocker, Thomas Siegert, and Laura Eisenberger
\end{center}

\vspace{2em}

\twocolumngrid

\section{Emission from cooling positrons}
\begin{figure}[h]
   \centering
   \includegraphics[width=1.0\linewidth]{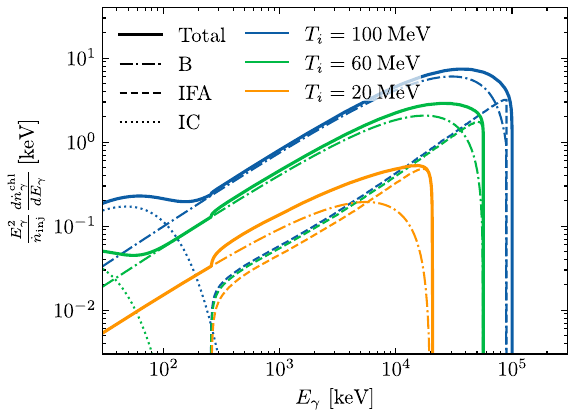}
   \caption{$\gamma$-ray emission (normalized to the injection rate) from cooling positrons, computed using \textsc{criptic}, in our benchmark ISM model (See appendix of main paper), for 3 different injection energies.
   Thick, solid: Total cooling emission; dot-dashed: B; dashed: IFA; dotted: IC from injected positrons.}
   \label{fig:dflux_chl}
\end{figure}

As discussed in the main text, the cooling radiation from positrons includes contributions from Bremsstrahlung (B), Inverse Compton (IC) and In-flight Annihilation (IFA). For electrons, only B and IC contribute to the cooling radiation. In our analysis, we calculate these components using the code \textsc{criptic} and present the spectral fit in Figure \ref{fig:bestfit-ptsrc-09} of the main paper. In Figure \ref{fig:dflux_chl} we present a breakup of these emission channels for relativistic positrons injected into the ISM. Note that this emission is sub-dominant to the thermalized radiation when the injected positrons are  mildly relativistic or non-relativistic.

\section{Varying the composition and density of the ISM}

As discussed in the main text, the exact shape of the ISM photon production function $\mathrm{d}n_\gamma^{\mathrm{cool}, e^\pm}/\mathrm{d}E_\gamma$ depends on the chemical state of the ISM (predominantly ionized versus atomic) and on the strengths of the interstellar magnetic and radiation fields, which in turn set the importance of synchrotron and inverse Compton losses to collisional losses, as parameterized by $f_\mathrm{sync}$ and $f_\mathrm{IC}$ \cite{Krumholz2023}. These conditions in turn depend on where in the Galaxy the positrons propagate and annihilate.

In order to understand how varying these parameters would affect our final conclusion, we experimented with the following sets of ISM conditions:
\begin{itemize}
    \item I: Ionized medium with $f_{\rm IC} = f_{\rm synch} = 10^{-5}$
    \item II: Ionized medium with $f_{\rm IC} = f_{\rm synch} = 10^{-6}$ (the benchmark model presented in the main text)
    \item III: Ionized medium with $f_{\rm IC} = f_{\rm synch} = 10^{-7}$
    \item IV: Ionized medium with $f_{\rm IC} = f_{\rm synch} = 10^{-8}$
    \item V: Atomic medium with $f_{\rm IC} = f_{\rm synch} = 10^{-6}$
\end{itemize}

In the full tables and figures below we present results for each of these ISM compositions, demonstrating that the qualitative conclusions do not change significantly between them.

\renewcommand{\arraystretch}{1.5}
\newcolumntype{P}[1]{>{\centering\arraybackslash}p{#1}}

\section{Comparing CGRO data scaling Models}

Again as discussed in the main text, we consider three models for how to scale the CGRO data relative to the INTEGRAL data for each ROI: \flatt, \likefive, and \ptsrc. We choose between these models based on the relative probabilities calculated from Equations \ref{eq:AIC}-\ref{eq:bayes} (of the main paper). In Table \ref{tab:bayes} we present the resulting relative probabilities, both for the Benchmark ISM model and the four alternatives described above. The results vary very little with ISM model; for all models we find that the \ptsrc~scaling yields the highest relative probability in small ROIs (5\degree~and 9\degree), even though the evidence is not very significant on Jeffreys' scale. The \flatt~model performs the best in larger ROI's (18\degree~and 95\degree), with a very strong to decisive significance on Jeffreys' scale.

A physical model that would fit both the behaviors of the CGRO emission would consist of an extended emission component over the sky along with a more localized component corresponding to the bulge. Such a distribution is qualitatively consistent with the maximum entropy reconstruction of the full-sky COMPTEL map from \cite{Strong19a}, which shows a strong emission peak coincident with the Galactic Center, which would likely be dominant in an observation with a small field of view, but also strong emission from the Galactic Plane that would begin to dominate for sufficiently large fields of view. While this qualitative consistency is comforting, it is challenging to do a more quantitative comparison, since the scaling model we have fit here is for the relative scaling between CGRO and INTEGRAL data rather than the absolute flux, and the morphology of the latter is also subject to uncertainties pertaining to the coded-mask nature of the instrument.

The choice of scaling model does affect the inferred upper limits on positron injection energy, because these constraints come primarily from the CGRO data. Thus scalings that reduce the CGRO data points relative to the SPI data also result in lower inferred maximum positron energies. Indeed, this effect is already visible in a close examination of Figure 2 of the main text, where the best-fitting model for the $9^\circ$ ROI places the CGRO data points slightly below what a straightforward extrapolation of the INTEGRAL data would suggest, and as a result the best-fitting model lies systematically below (but still well within the error bars of) the INTEGRAL data points. Scaling models that lowered the CGRO point even further relative to the INTEGRAL ones would exaggerate this effect, and force the fits toward lower positron injection energies. However,  these models are also strongly disfavored by the AIC because the large discontinuities in the data between the INTEGRAL and CGRO data that they introduce make the overall fit much worse. 

\begin{table*}
    \centering
    \begin{tabular}{ P{3cm} P{1.8cm} P{2cm} P{2cm} P{2cm} }
\hline
ISM conditions           & ROI & \flatt & \likefive & \ptsrc \\ \hline

\multirow{4}{*}{\begin{tabular}[c]{@{}c@{}}I. Ionized\\ $f_{\rm IC} = 10^{-5}$\end{tabular}} 
                         & 5\degree circle  & 0.2188 &  0.3341 & \textbf{0.4471} \\ 
                         & 9\degree circle  & 0.2012 &  0.354 & \textbf{0.4448}  \\ 
                         & 18\degree circle & \textbf{0.9748} &  0.0223 & 0.0029 \\  
                         & 95\degree square & \textbf{1.0} &  $5.131\times 10^{-10}$ & $2.671\times 10^{-13}$ \\ \hline

\multirow{4}{*}{\begin{tabular}[c]{@{}c@{}}II. Ionized\\ $f_{\rm IC} = 10^{-6}$ \\ Benchmark model\end{tabular}} 
                         & 5\degree circle  & 0.2383 &  0.3374 & \textbf{0.4243} \\  
                         & 9\degree circle  & 0.1738 &  0.385 & \textbf{0.4412} \\ 
                         & 18\degree circle & \textbf{0.9723} &  0.0254 & 0.0022 \\  
                         & 95\degree square & \textbf{1.0} &  $3.205\times10^{-10}$ & $3.685\times 10^{-13}$ \\ \hline

\multirow{4}{*}{\begin{tabular}[c]{@{}c@{}}III. Ionized\\ $f_{\rm IC} = 10^{-7}$\end{tabular}}
                         & 5\degree circle  & 0.185 &  0.3585 & \textbf{0.4565} \\  
                         & 9\degree circle  & 0.2106 &  0.2779 & \textbf{0.5114} \\  
                         & 18\degree circle & \textbf{0.969} &  0.0277 & 0.0034 \\ 
                         & 95\degree square & \textbf{1.0} &  $4.277\times 10^{-10}$ & $3.058\times10^{-13}$ \\ \hline
                       
\multirow{4}{*}{\begin{tabular}[c]{@{}c@{}}IV. Ionized\\ $f_{\rm IC} = 10^{-8}$\end{tabular}} 
                         & 5\degree circle  & 0.2176 &  0.3407 & \textbf{0.4418}  \\ 
                         & 9\degree circle  & 0.1949 &  0.3061 & \textbf{0.4991}  \\ 
                         & 18\degree circle & \textbf{0.9759} &  0.0215 & 0.0026  \\ 
                         & 95\degree square & \textbf{1.0} &  $4.328\times 10^{-10}$ & $3.358\times 10^{-13}$  \\ \hline

\multirow{4}{*}{\begin{tabular}[c]{@{}c@{}}V. Atomic\\ $f_{\rm IC} = 10^{-6}$\end{tabular}} 
                         & 5\degree circle  & 0.3007 &  0.3132 & \textbf{0.3861}   \\ 
                         & 9\degree circle  & 0.2428 &  0.3520 & \textbf{0.4052}   \\  
                         & 18\degree circle & \textbf{0.9293} &  0.0586 & 0.0121 \\  
                         & 95\degree square & \textbf{1.0} &  $3.865\times 10^{-9}$ & $1.403\times 10^{-11}$ \\ \hline
            
\end{tabular}
    \caption{Comparison of relative probabilities for each ROI under various ISM conditions}
    \label{tab:bayes}
\end{table*}

\section{Limits in the Absence of CGRO data}

\begin{figure}[ht]
    \centering
    \includegraphics[width=1.0\linewidth]{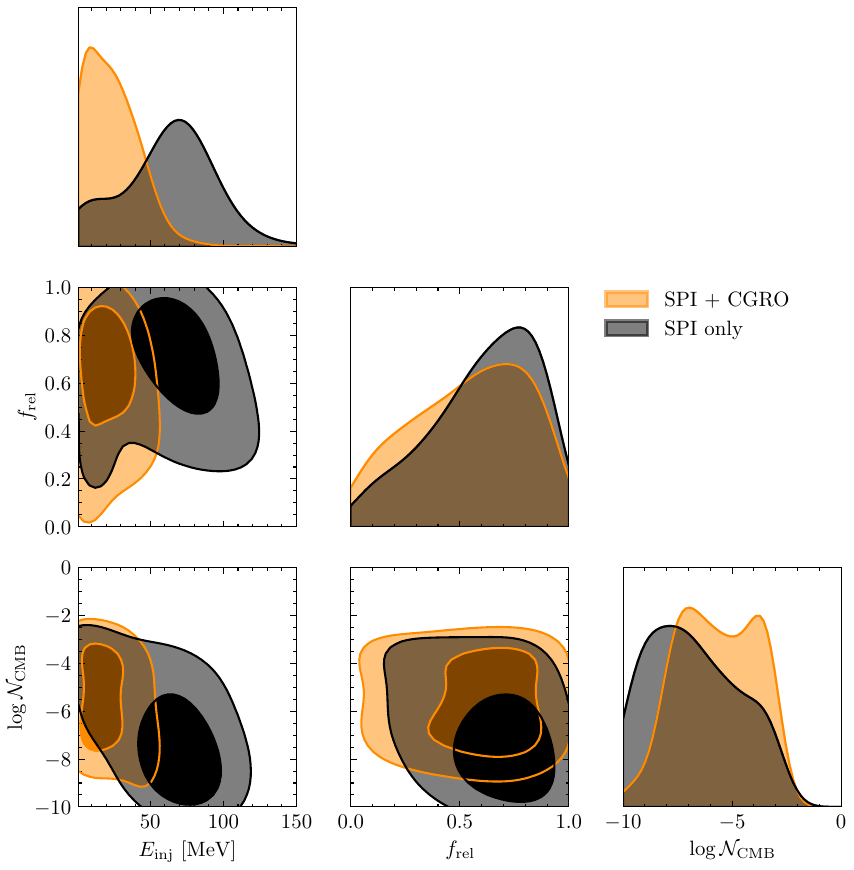}
    \caption{Comparison of posteriors for parameters $E_{\rm inj}$, $f_{\rm rel}$ and $\mathcal{N}_{\rm CMB}$ for our fiducial choice of data: SPI + CGRO (orange) vs. using only SPI data (black). We find that the latter prefers higher injection energies and higher contribution of relativistic positrons.}
    \label{fig:compare-spi-cgro}
\end{figure}

\begin{figure}[ht]
    \centering
    \includegraphics[width=1.0\linewidth]{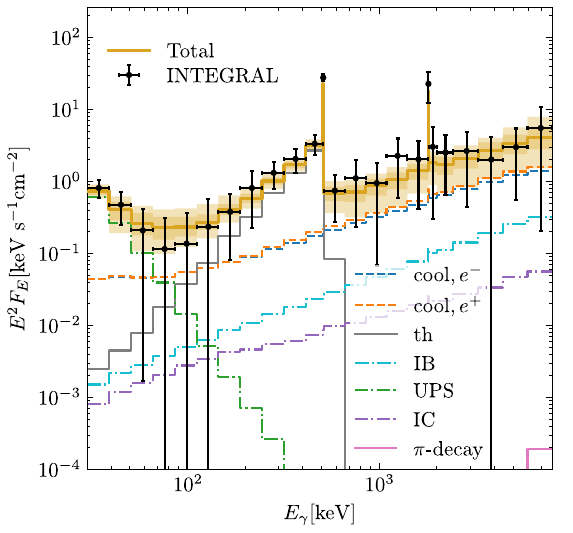}
    \caption{Spectral fit of INTEGRAL/SPI data sampled from posterior of SPI-only model fits. Yellow bands show median, $1\sigma$ and $2\sigma$ bands from the posterior, while other lines show median contributions from single emission channels similar to Figure \ref{fig:bestfit-ptsrc-09} of main paper.}
    \label{fig:spectrum-spi-only}
\end{figure}

We present a comparison of posteriors for our fiducial case against a situation where only INTEGRAL/SPI fluxes are used to constrain our model, without using the gamma-ray flux at energies higher than $\sim$ 10 MeV. Owing to a significant gap in fluxes below 10 MeV as measured by INTEGRAL and above it as measured by COMPTEL, the limits on injection energy obtained in the absence of these COMPTEL data is large. Using only INTEGRAL fluxes, the 95\% C.L. upper limit on the injection energy turns out to be 102 MeV, which would be a greater relaxation on the injection energy as compared to our fiducial analysis. This highlights the need for flux measurements at these energies in order to efficiently constrain positron injection energies.
\\\\
We have shown a comparison of selected model parameters for the two cases in Figure \ref{fig:compare-spi-cgro}. The resulting spectral fit from only SPI data is displayed in Figure \ref{fig:spectrum-spi-only}.

\section{Posteriors on full model parameters}

\begin{figure*}[!ht]
    \centering
    \begin{subfigure}{0.45\textwidth}
        \includegraphics[width=0.8\textwidth]{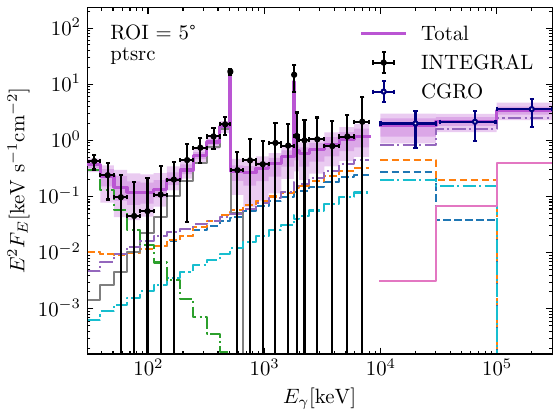}
    \end{subfigure}
    \begin{subfigure}{0.45\textwidth}
        \includegraphics[width=0.8\textwidth]{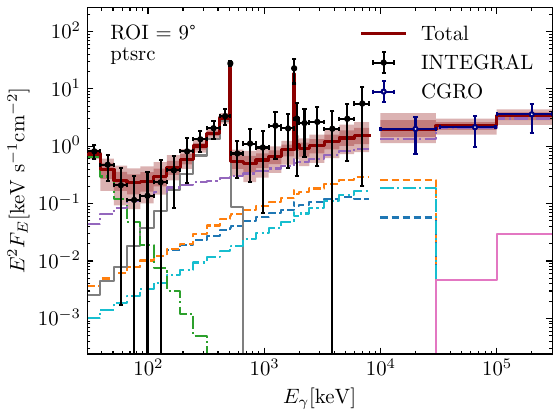}
    \end{subfigure}
    \begin{subfigure}{0.45\textwidth}
        \includegraphics[width=0.8\textwidth]{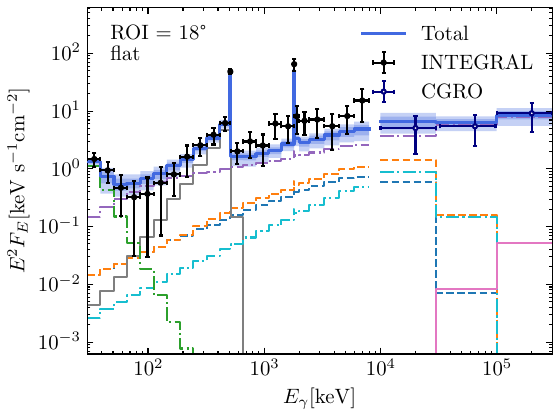}
    \end{subfigure}
    \begin{subfigure}{0.45\textwidth}
        \includegraphics[width=0.8\textwidth]{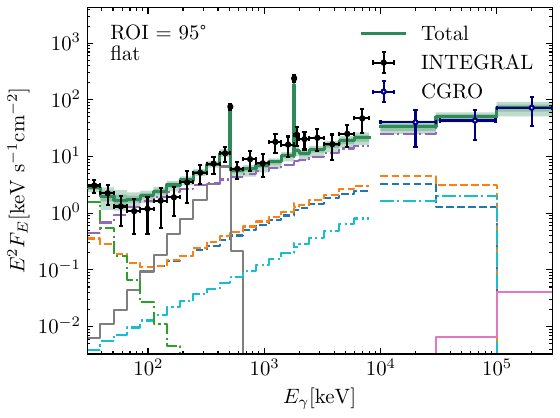}
    \end{subfigure}
    \caption{Median spectral fits for each region of interest (using the scaling model with highest relative probability), similar to Figure \ref{fig:bestfit-ptsrc-09} of main paper, for an ionized medium with $f_{\rm IC}=10^{-5}$ (ISM model I). The individual components of the $\gamma$-ray emission are labeled the same way as in Fig. \ref{fig:bestfit-ptsrc-09} of the main paper.}
    \label{fig:spectral_fits_ionized_5.0}
\end{figure*}

\begin{figure*}[!ht]
    \centering
    \begin{subfigure}{0.45\textwidth}
        \includegraphics[width=0.8\textwidth]{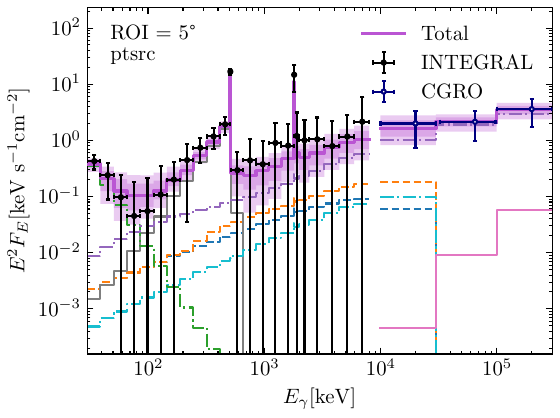}
    \end{subfigure}
    \begin{subfigure}{0.45\textwidth}
        \includegraphics[width=0.8\textwidth]{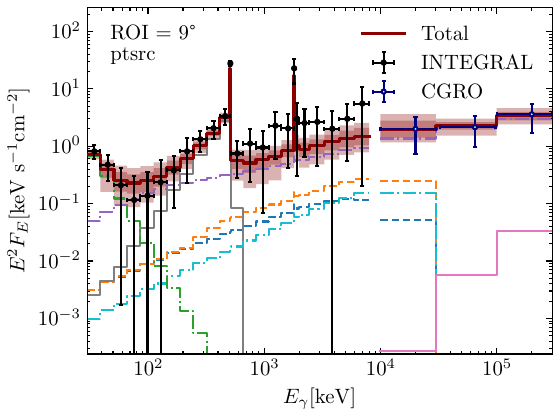}
    \end{subfigure}
    \begin{subfigure}{0.45\textwidth}
        \includegraphics[width=0.8\textwidth]{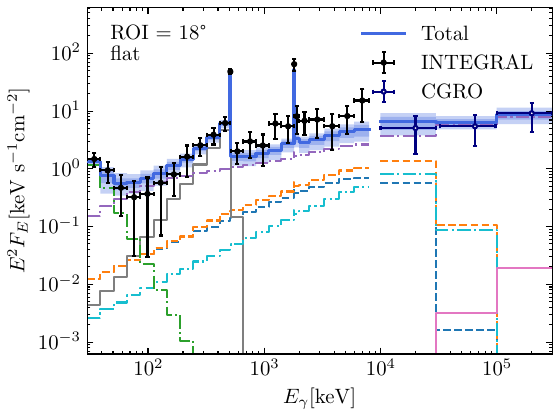}
    \end{subfigure}
    \begin{subfigure}{0.45\textwidth}
        \includegraphics[width=0.8\textwidth]{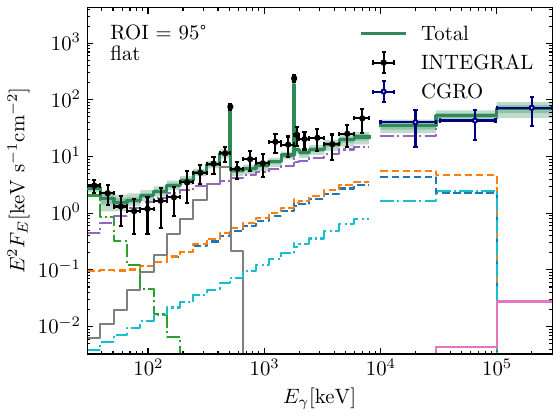}
    \end{subfigure}
    \caption{Same as \ref{fig:spectral_fits_ionized_5.0}, for the fiducial ISM model (II).}
    \label{fig:spectral_fits_ionized_6.0}
\end{figure*}

\begin{figure*}[!ht]
    \centering
    \begin{subfigure}{0.45\textwidth}
        \includegraphics[width=0.8\textwidth]{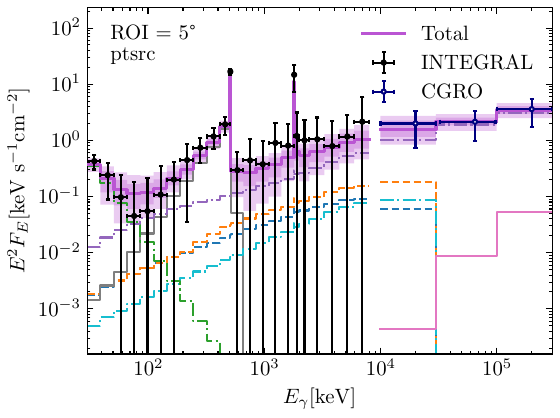}
    \end{subfigure}
    \begin{subfigure}{0.45\textwidth}
        \includegraphics[width=0.8\textwidth]{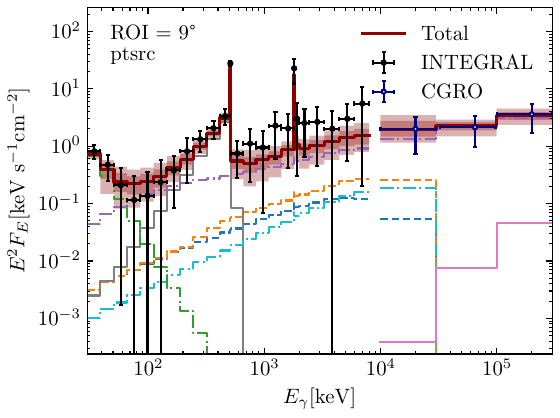}
    \end{subfigure}
    \begin{subfigure}{0.45\textwidth}
        \includegraphics[width=0.8\textwidth]{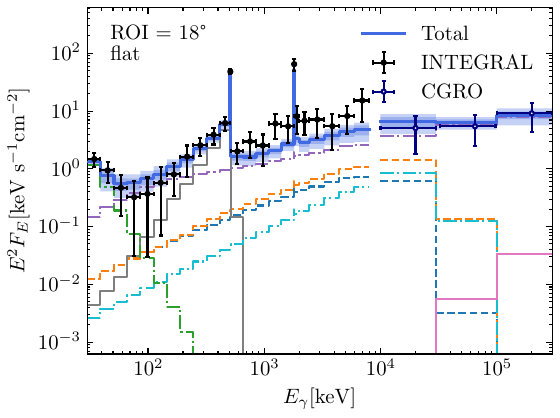}
    \end{subfigure}
    \begin{subfigure}{0.45\textwidth}
        \includegraphics[width=0.8\textwidth]{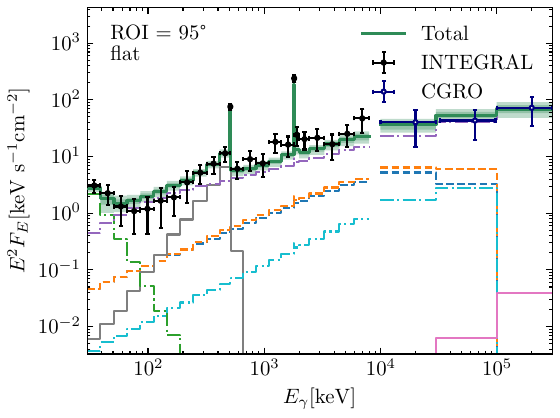}
    \end{subfigure}
    \caption{Same as \ref{fig:spectral_fits_ionized_5.0}, for an ionized medium with $f_{\rm IC}=10^{-7}$ (III)}
    \label{fig:spectral_fits_ionized_7.0}
\end{figure*}

\begin{figure*}[!ht]
    \centering
    \begin{subfigure}{0.45\textwidth}
        \includegraphics[width=0.8\textwidth]{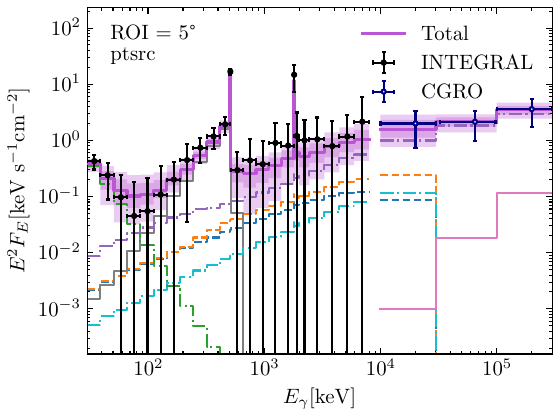}
    \end{subfigure}
    \begin{subfigure}{0.45\textwidth}
        \includegraphics[width=0.8\textwidth]{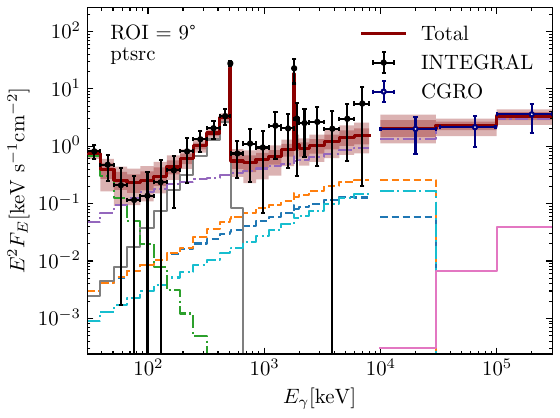}
    \end{subfigure}
    \begin{subfigure}{0.45\textwidth}
        \includegraphics[width=0.8\textwidth]{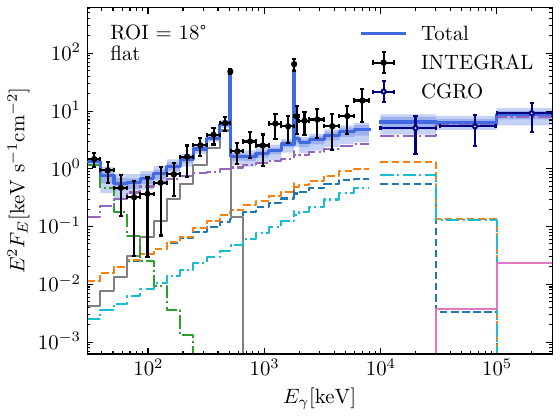}
    \end{subfigure}
    \begin{subfigure}{0.45\textwidth}
        \includegraphics[width=0.8\textwidth]{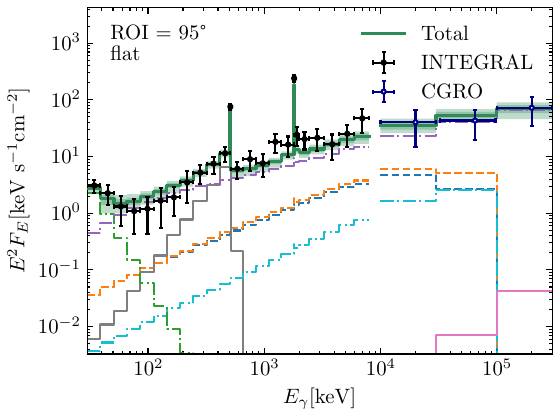}
    \end{subfigure}
    \caption{Same as \ref{fig:spectral_fits_ionized_5.0}, for an ionized medium with $f_{\rm IC}=10^{-8}$ (IV)}
    \label{fig:spectral_fits_ionized_8.0}
\end{figure*}

\begin{figure*}[!ht]
    \centering
    \begin{subfigure}{0.45\textwidth}
        \includegraphics[width=0.8\textwidth]{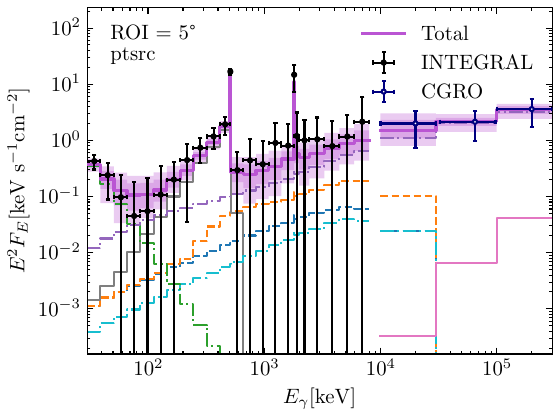}
    \end{subfigure}
    \begin{subfigure}{0.45\textwidth}
        \includegraphics[width=0.8\textwidth]{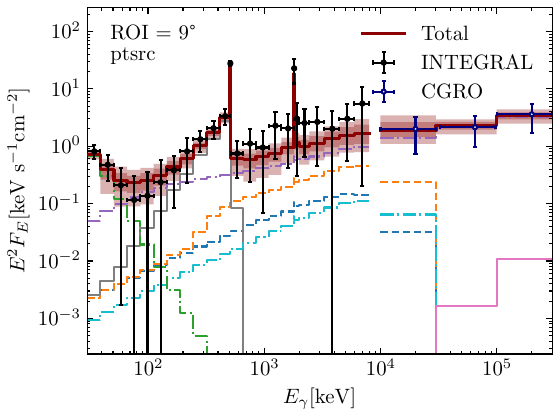}
    \end{subfigure}
    \begin{subfigure}{0.45\textwidth}
        \includegraphics[width=0.8\textwidth]{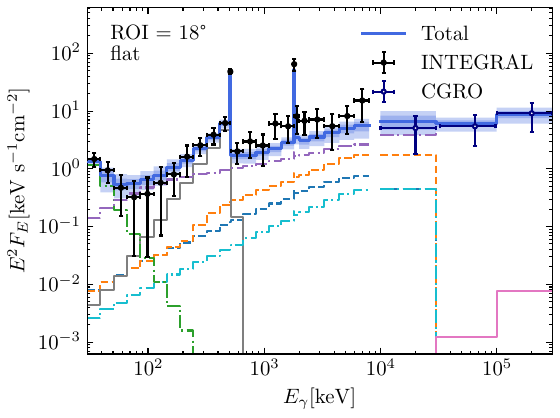}
    \end{subfigure}
    \begin{subfigure}{0.45\textwidth}
        \includegraphics[width=0.8\textwidth]{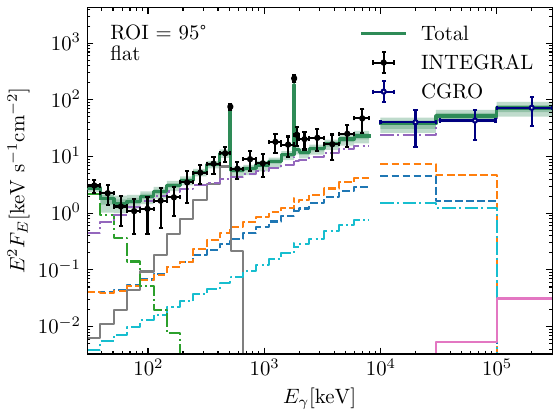}
    \end{subfigure}
    \caption{Same as \ref{fig:spectral_fits_ionized_5.0}, for an atomic medium with $f_{\rm IC}=10^{-6}$ (V)}
    \label{fig:spectral_fits_atomic_6.0}
\end{figure*}

\begin{figure*}[!ht]
    \centering
    \includegraphics[width=1.0\linewidth]{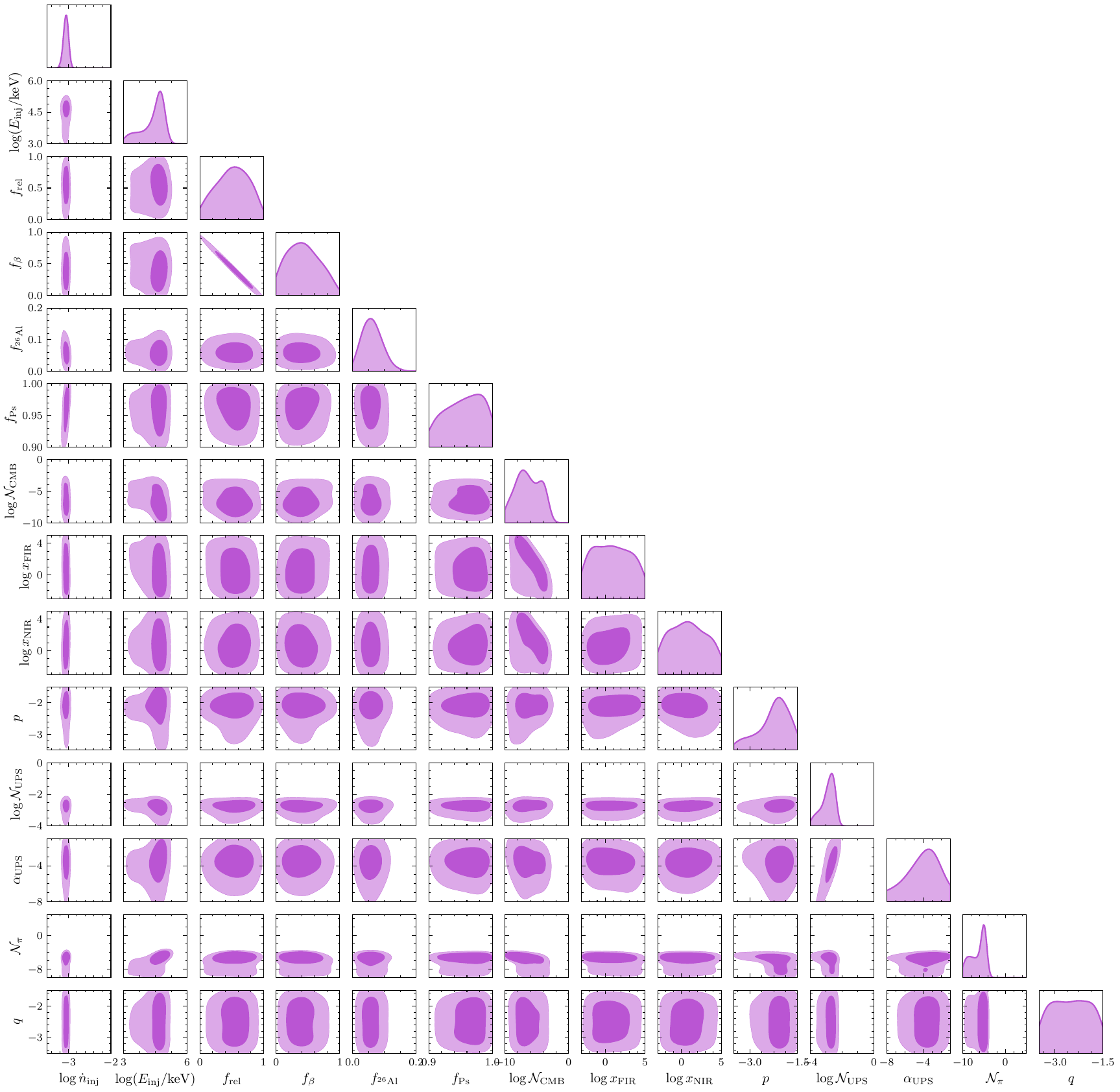}
    \caption{Full posteriors for (I) ionized medium, $f_{\rm IC}=10^{-5}$, for the 5\degree~ROI. We use the convention of Table \ref{tab:priors} for the units and normalization of each of these parameters. $E_{\rm inj}$ is in units of keV.}
    \label{fig:fullposterior_ionized_5.0_05}
\end{figure*}

\begin{figure*}[!ht]
    \centering
    \includegraphics[width=1.0\linewidth]{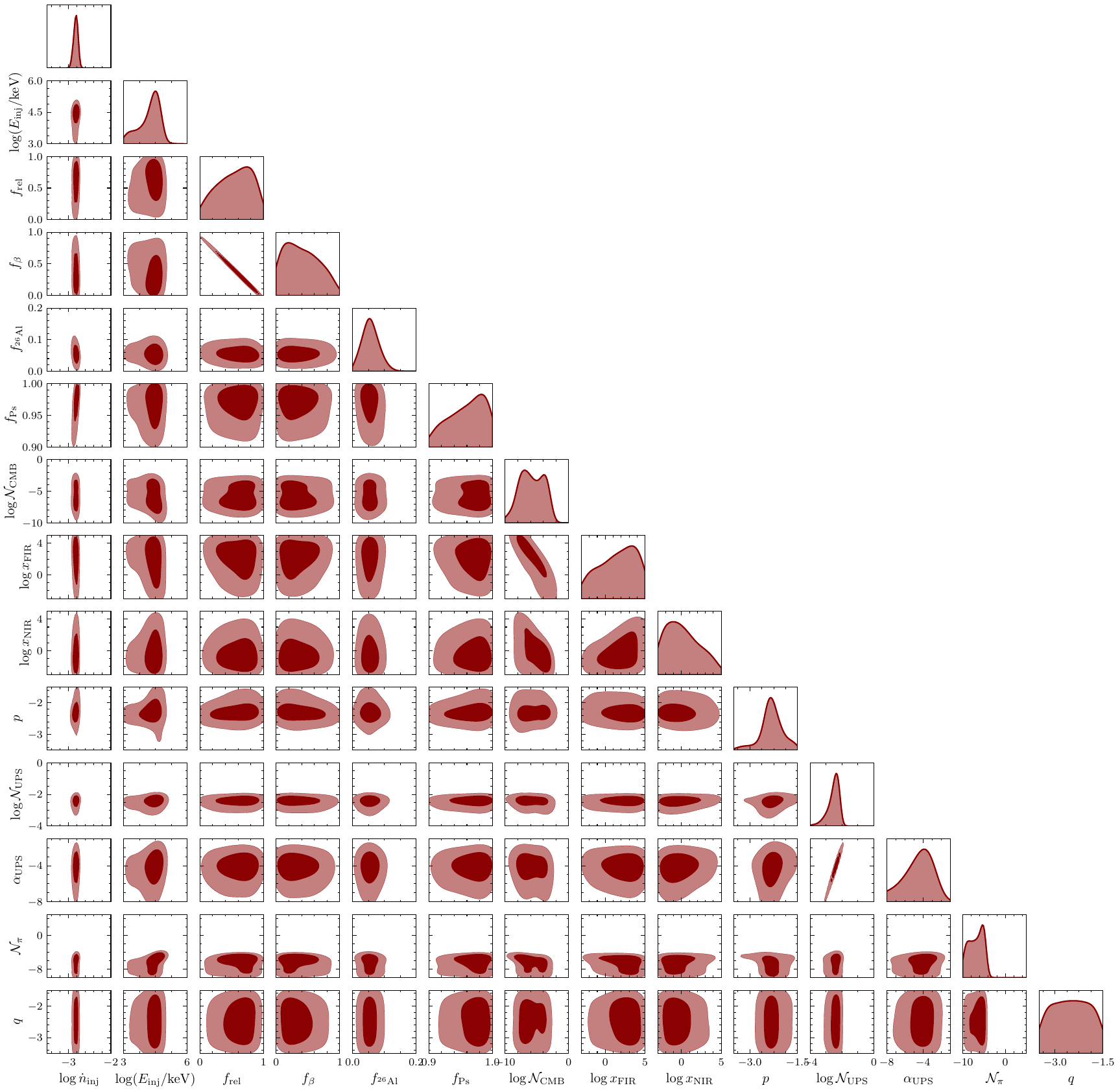}
    \caption{Full posteriors for (I) ionized medium, $f_{\rm IC}=10^{-5}$, for the 9\degree~ROI.}
    \label{fig:fullposterior_ionized_5.0_09}
\end{figure*}

\begin{figure*}[!ht]
    \centering
    \includegraphics[width=1.0\linewidth]{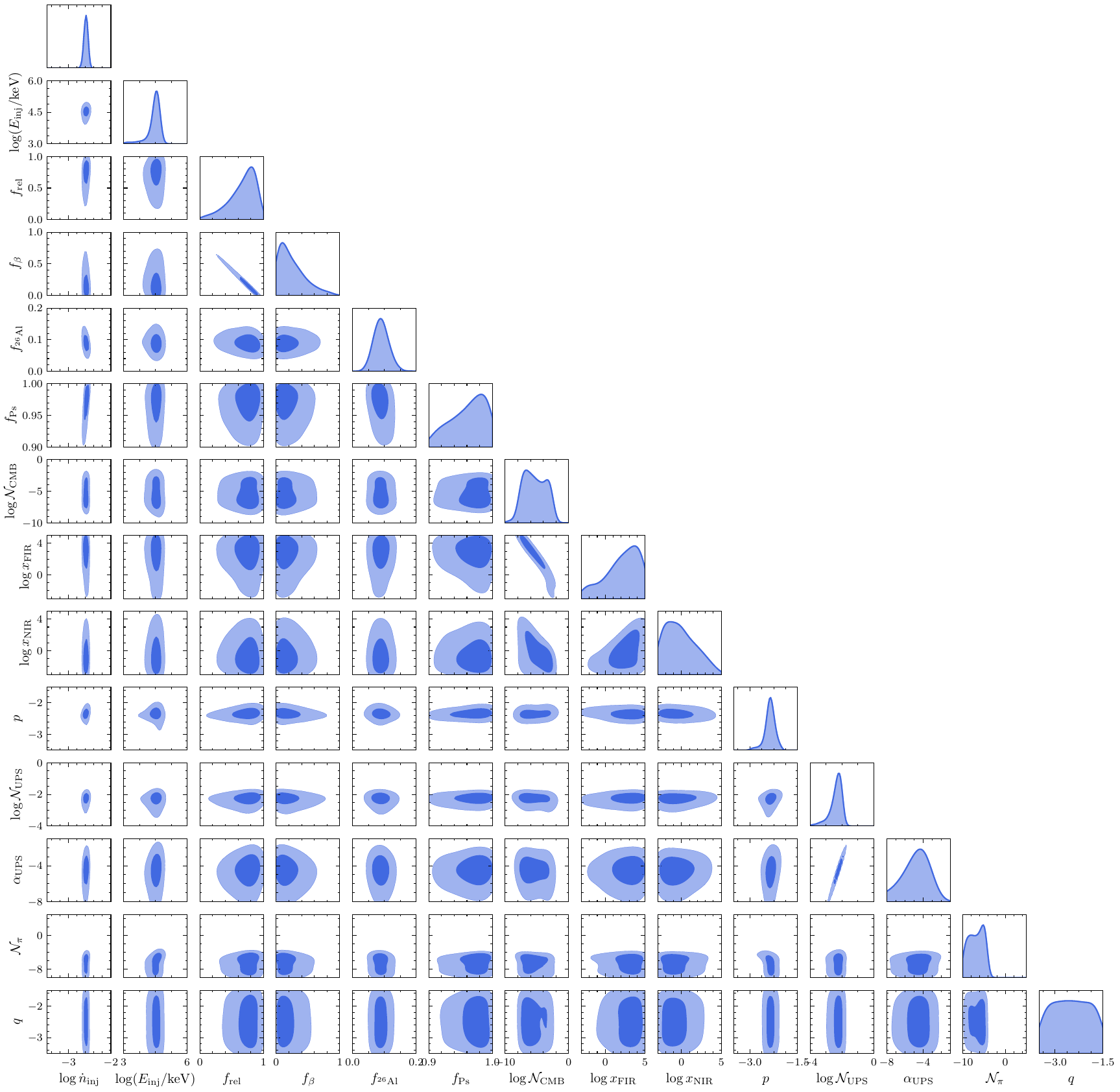}
    \caption{Full posteriors for (I) ionized medium, $f_{\rm IC}=10^{-5}$, for the 18\degree~ROI.}
    \label{fig:fullposterior_ionized_5.0_18}
\end{figure*}

\begin{figure*}[!ht]
    \centering
    \includegraphics[width=1.0\linewidth]{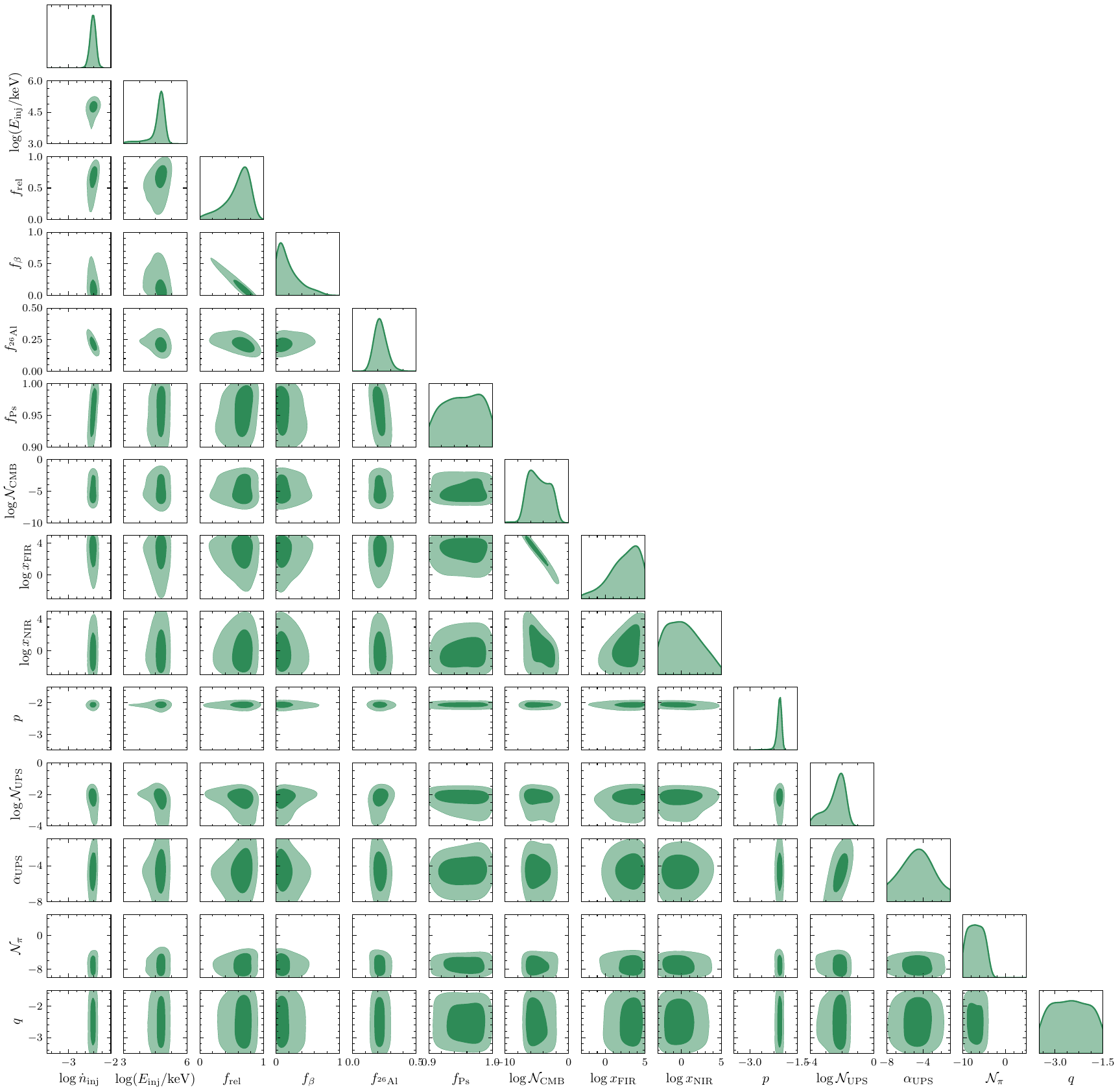}
    \caption{Full posteriors for (I) ionized medium, $f_{\rm IC}=10^{-5}$, for the 95\degree~ROI.}
    \label{fig:fullposterior_ionized_5.0_95}
\end{figure*}

\begin{figure*}[!ht]
    \centering
    \includegraphics[width=1.0\linewidth]{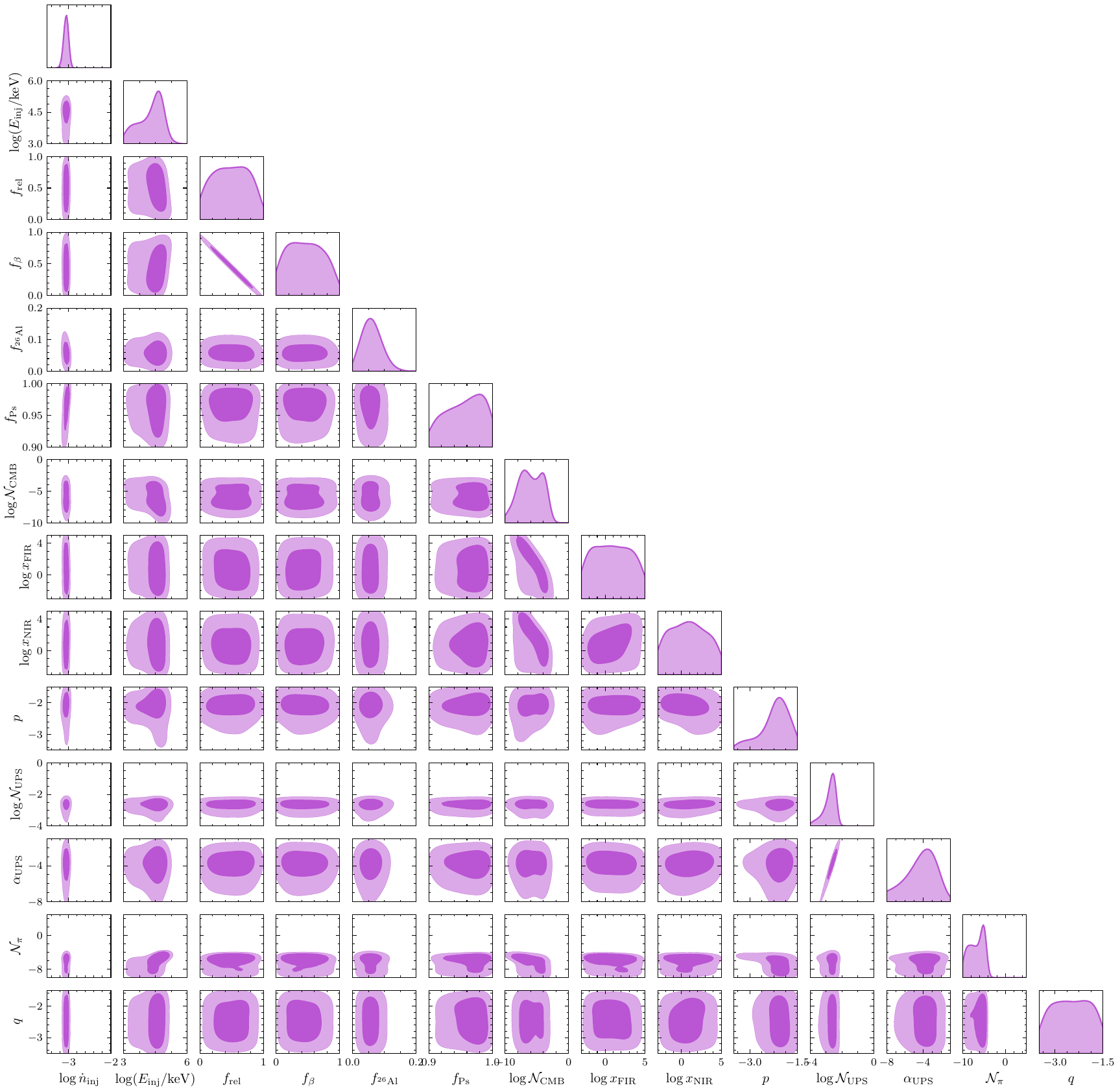}
    \caption{Full posteriors for (II; benchmark ISM model) ionized medium, $f_{\rm IC}=10^{-6}$, for the 5\degree~ROI.}
    \label{fig:fullposterior_ionized_6.0_05}
\end{figure*}

\begin{figure*}[!ht]
    \centering
    \includegraphics[width=1.0\linewidth]{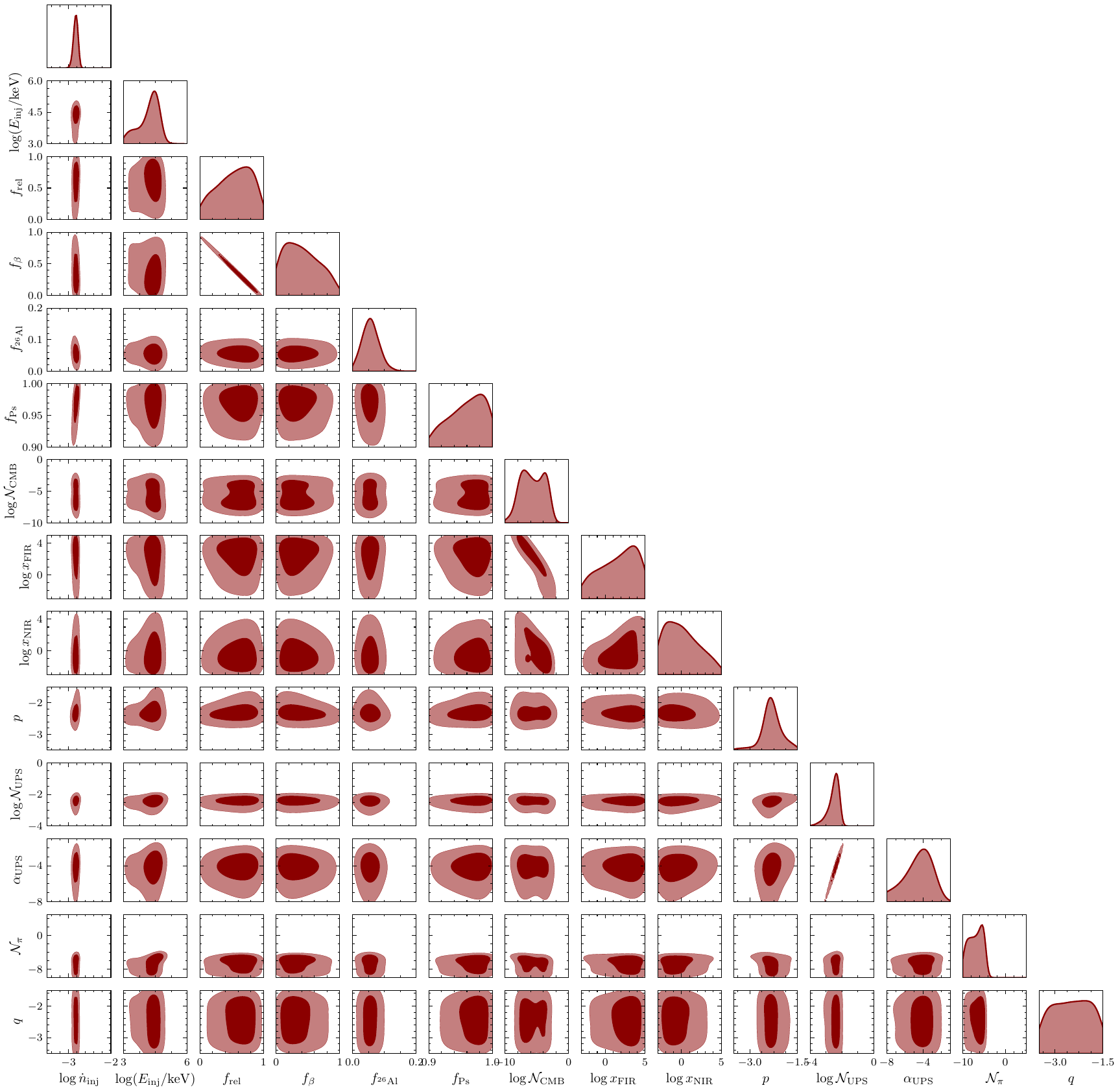}
    \caption{Full posteriors for (II; benchmark ISM model) ionized medium, $f_{\rm IC}=10^{-6}$, for the 9\degree~ROI.}
    \label{fig:fullposterior_ionized_6.0_09}
\end{figure*}

\begin{figure*}[!ht]
    \centering
    \includegraphics[width=1.0\linewidth]{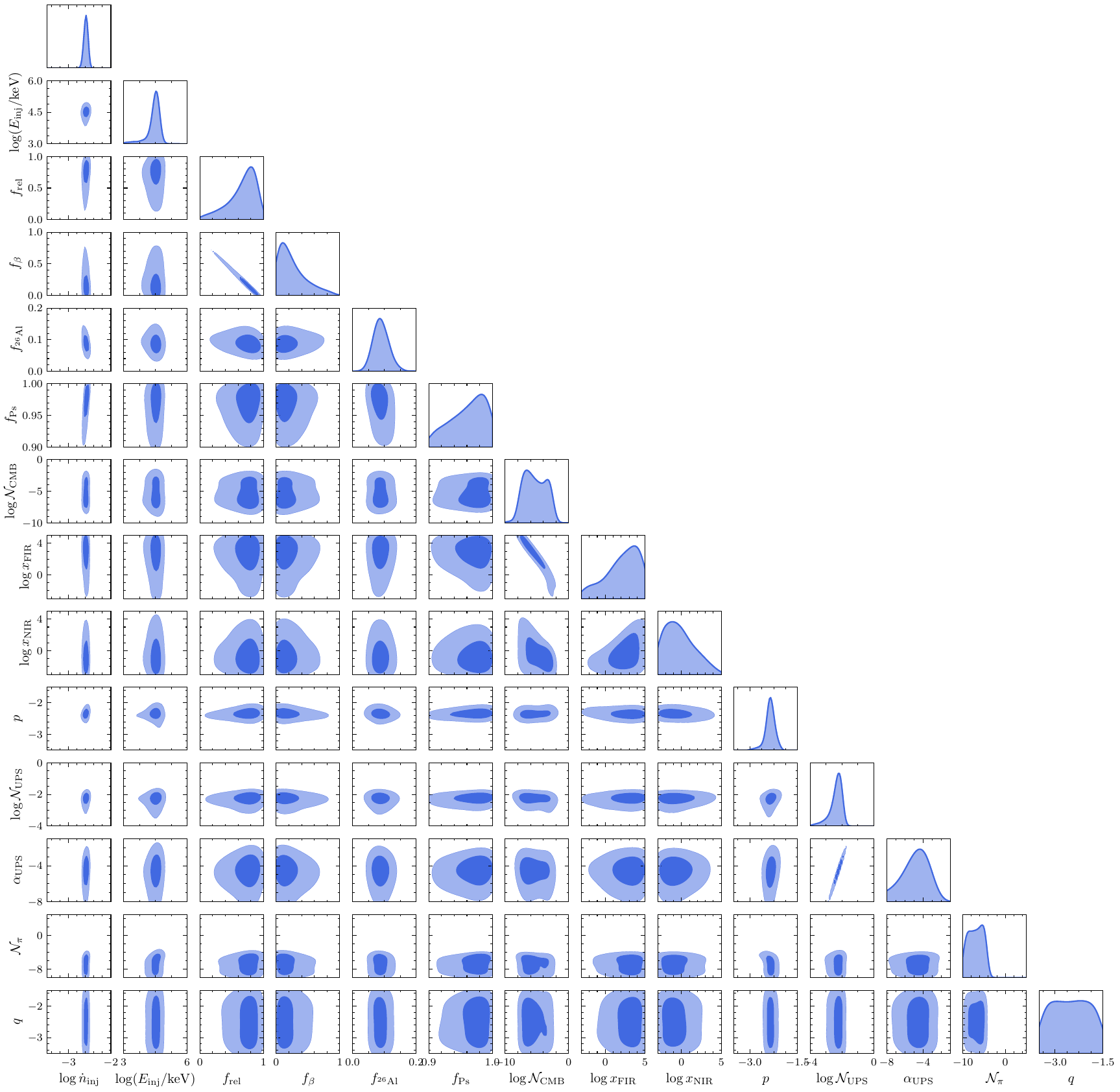}
    \caption{Full posteriors for (II; benchmark ISM model) ionized medium, $f_{\rm IC}=10^{-6}$, for the 18\degree~ROI.}
    \label{fig:fullposterior_ionized_6.0_18}
\end{figure*}

\begin{figure*}[!ht]
    \centering
    \includegraphics[width=1.0\linewidth]{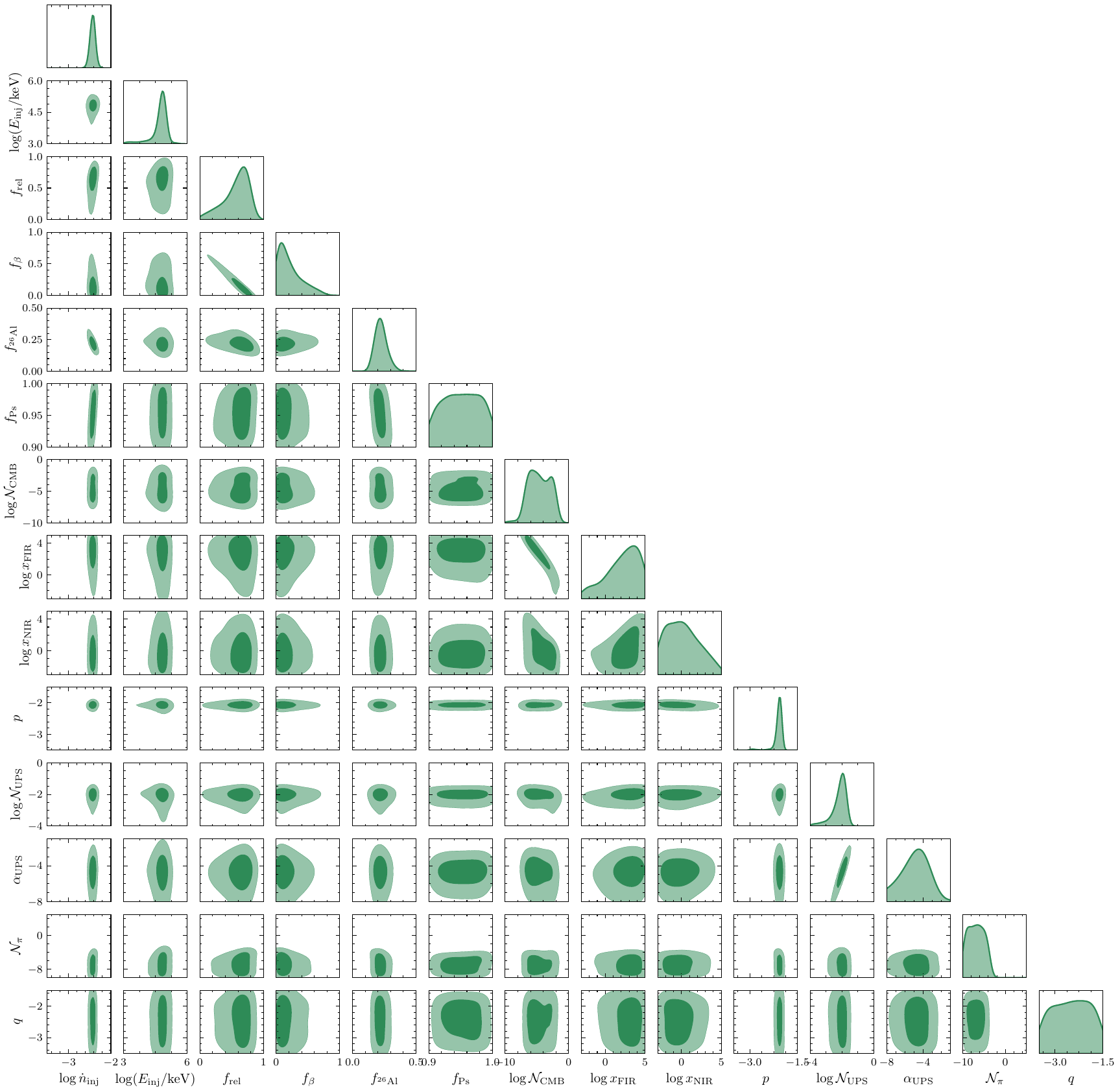}
    \caption{Full posteriors for (II; benchmark ISM model) ionized medium, $f_{\rm IC}=10^{-6}$, for the 95\degree~ROI.}
    \label{fig:fullposterior_ionized_6.0_95}
\end{figure*}

\begin{figure*}[!ht]
    \centering
    \includegraphics[width=1.0\linewidth]{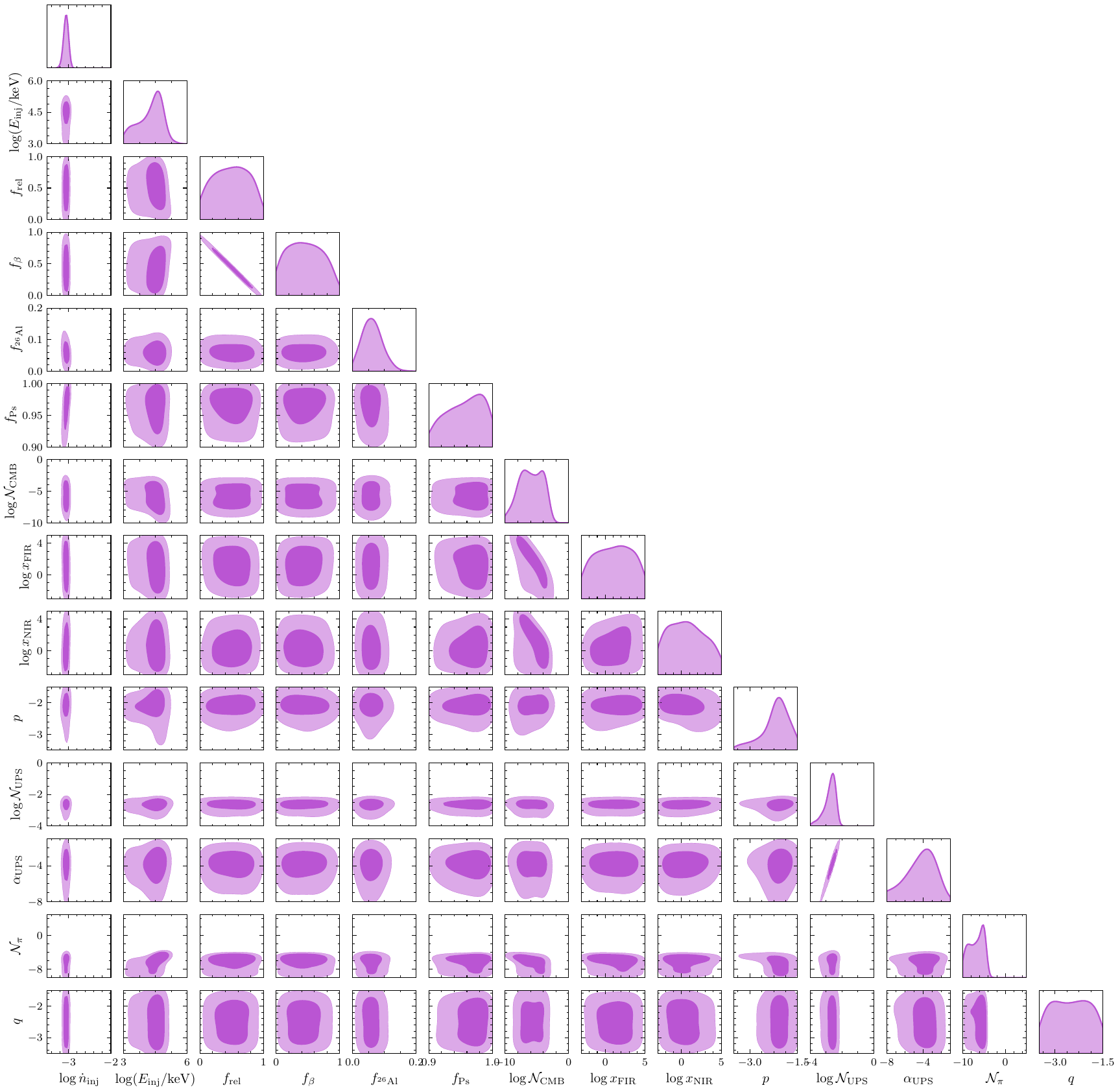}
    \caption{Full posteriors for (III) ionized medium, $f_{\rm IC}=10^{-7}$, for the 5\degree~ROI.}
    \label{fig:fullposterior_ionized_7.0_05}
\end{figure*}

\begin{figure*}[!ht]
    \centering
    \includegraphics[width=1.0\linewidth]{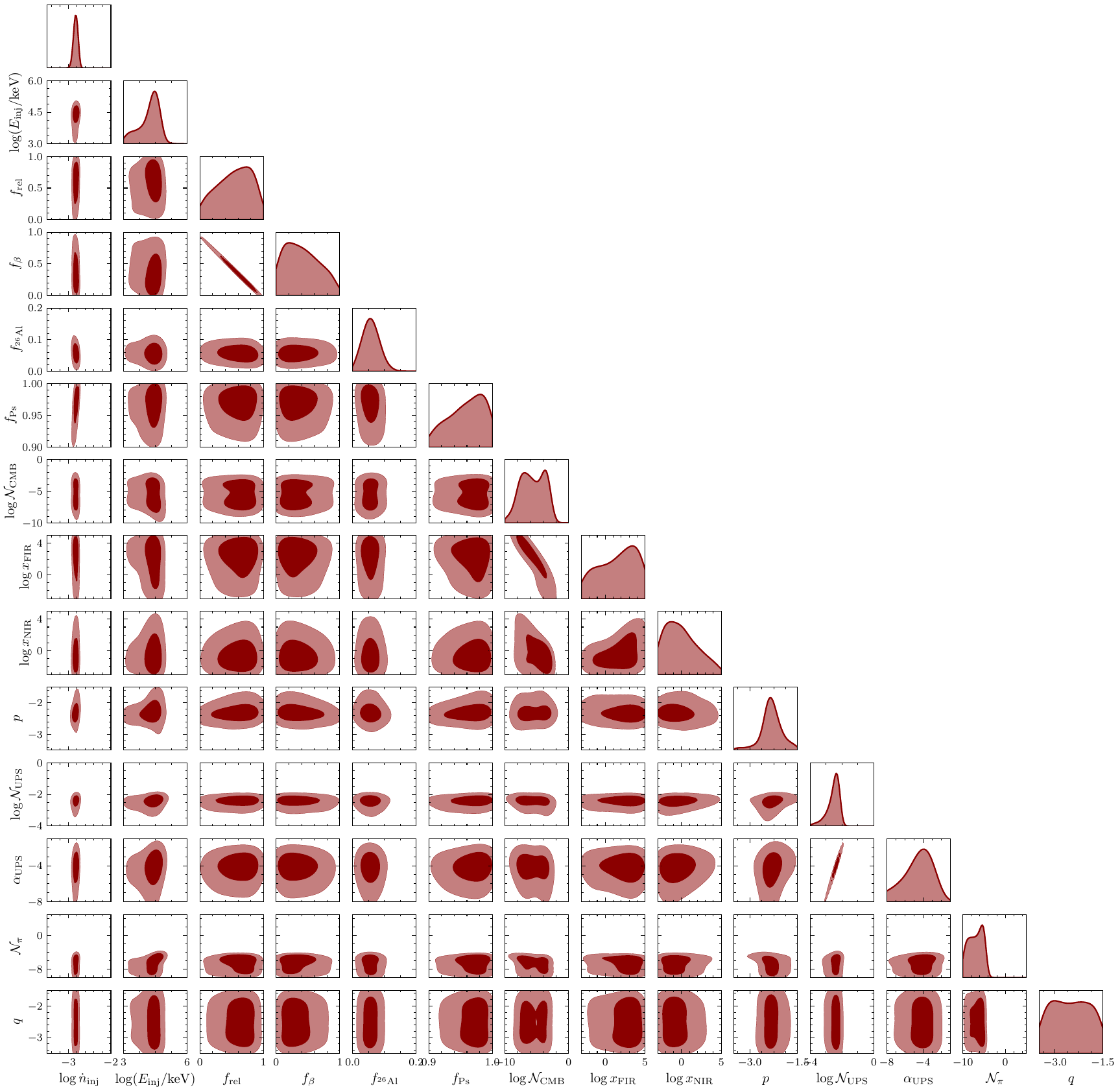}
    \caption{Full posteriors for (III) ionized medium, $f_{\rm IC}=10^{-7}$, for the 9\degree~ROI.}
    \label{fig:fullposterior_ionized_7.0_09}
\end{figure*}

\begin{figure*}[!ht]
    \centering
    \includegraphics[width=1.0\linewidth]{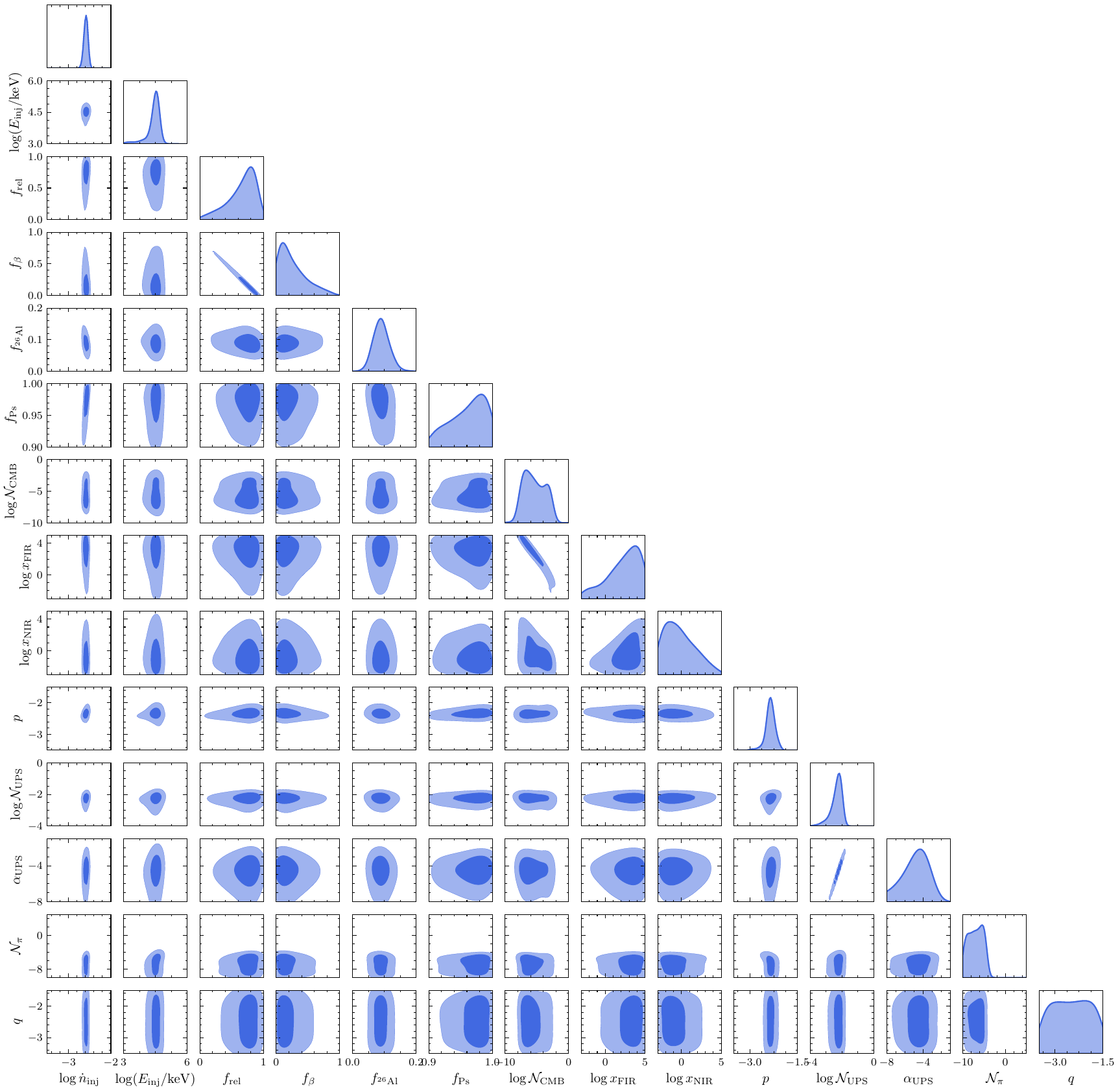}
    \caption{Full posteriors for (III) ionized medium, $f_{\rm IC}=10^{-7}$, for the 18\degree~ROI.}
    \label{fig:fullposterior_ionized_7.0_18}
\end{figure*}

\begin{figure*}[!ht]
    \centering
    \includegraphics[width=1.0\linewidth]{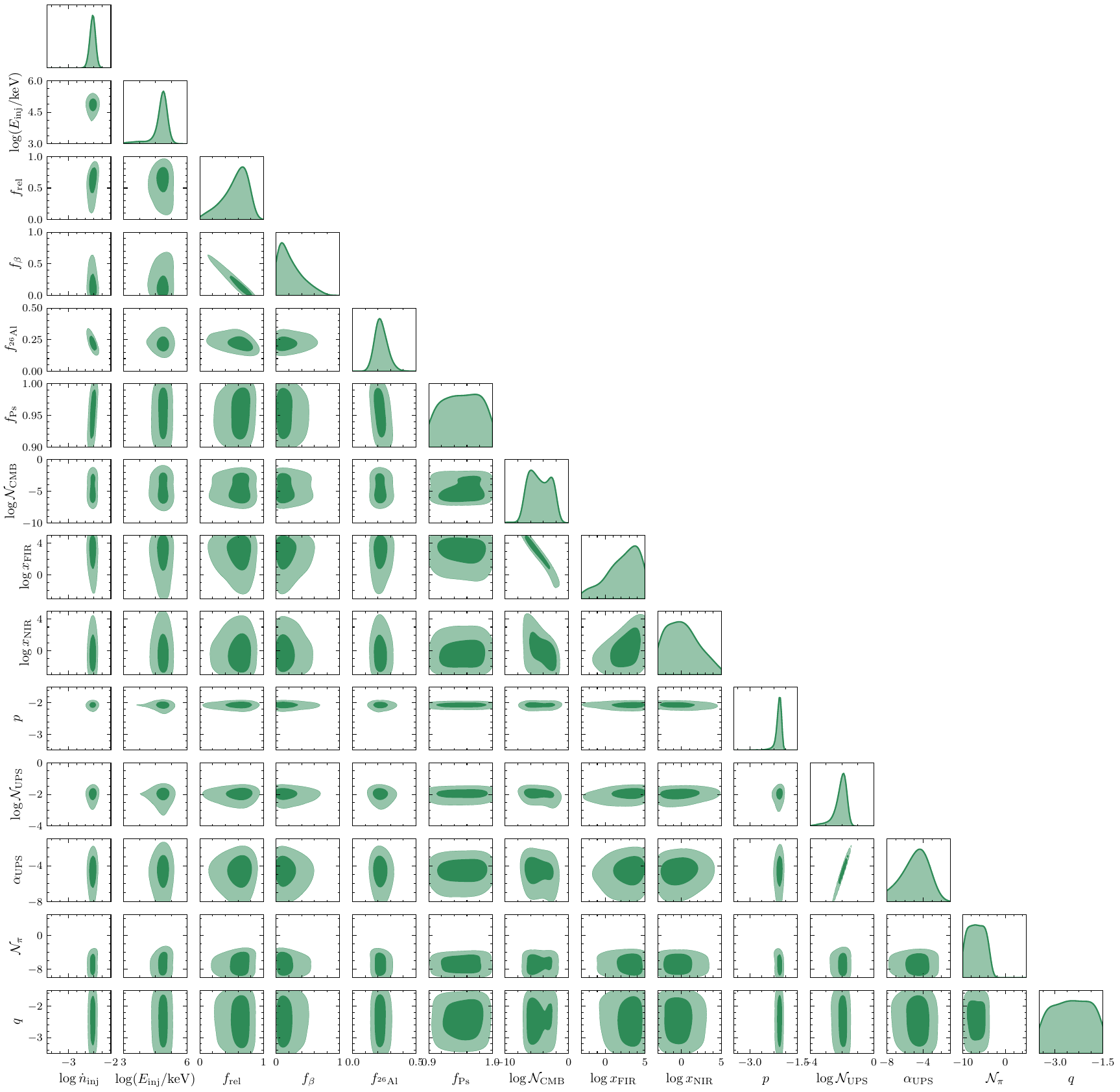}
    \caption{Full posteriors for (III) ionized medium, $f_{\rm IC}=10^{-7}$, for the 95\degree~ROI.}
    \label{fig:fullposterior_ionized_7.0_95}
\end{figure*}

\begin{figure*}[!ht]
    \centering
    \includegraphics[width=1.0\linewidth]{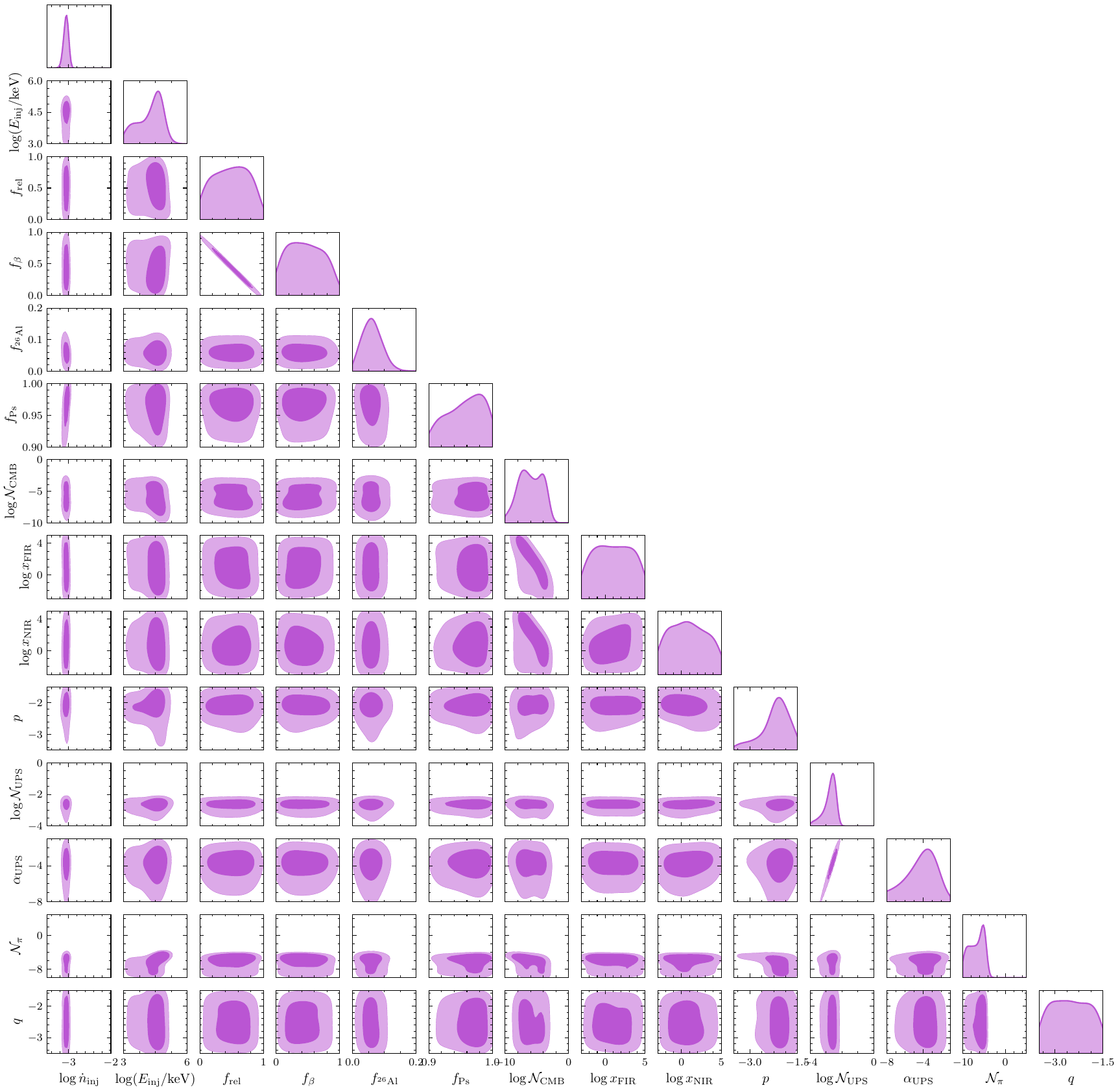}
    \caption{Full posteriors for (IV) ionized medium, $f_{\rm IC}=10^{-8}$, for the 5\degree~ROI.}
    \label{fig:fullposterior_ionized_8.0_05}
\end{figure*}

\begin{figure*}[!ht]
    \centering
    \includegraphics[width=1.0\linewidth]{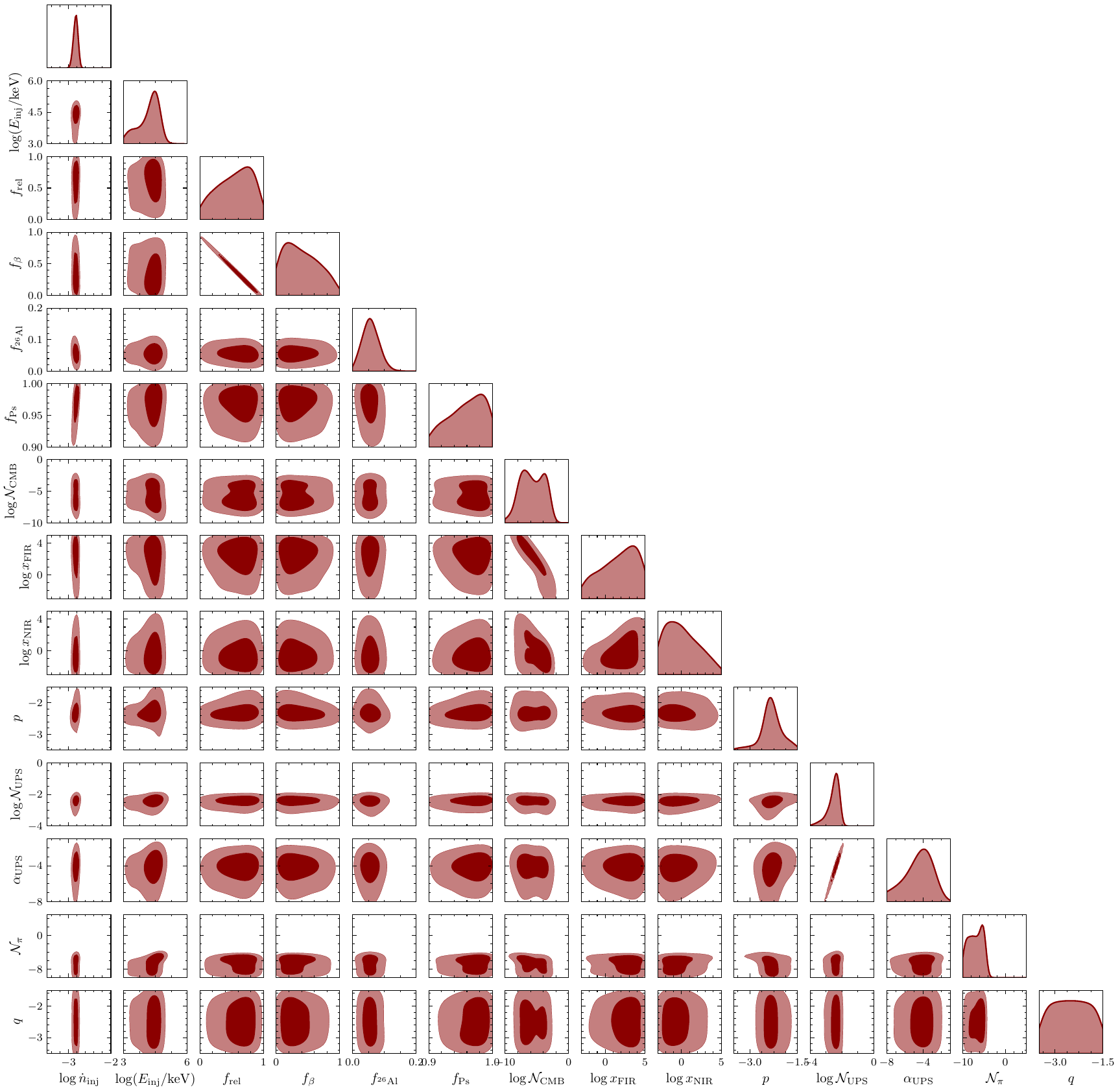}
    \caption{Full posteriors for (IV) ionized medium, $f_{\rm IC}=10^{-8}$, for the 9\degree~ROI.}
    \label{fig:fullposterior_ionized_8.0_09}
\end{figure*}

\begin{figure*}[!ht]
    \centering
    \includegraphics[width=1.0\linewidth]{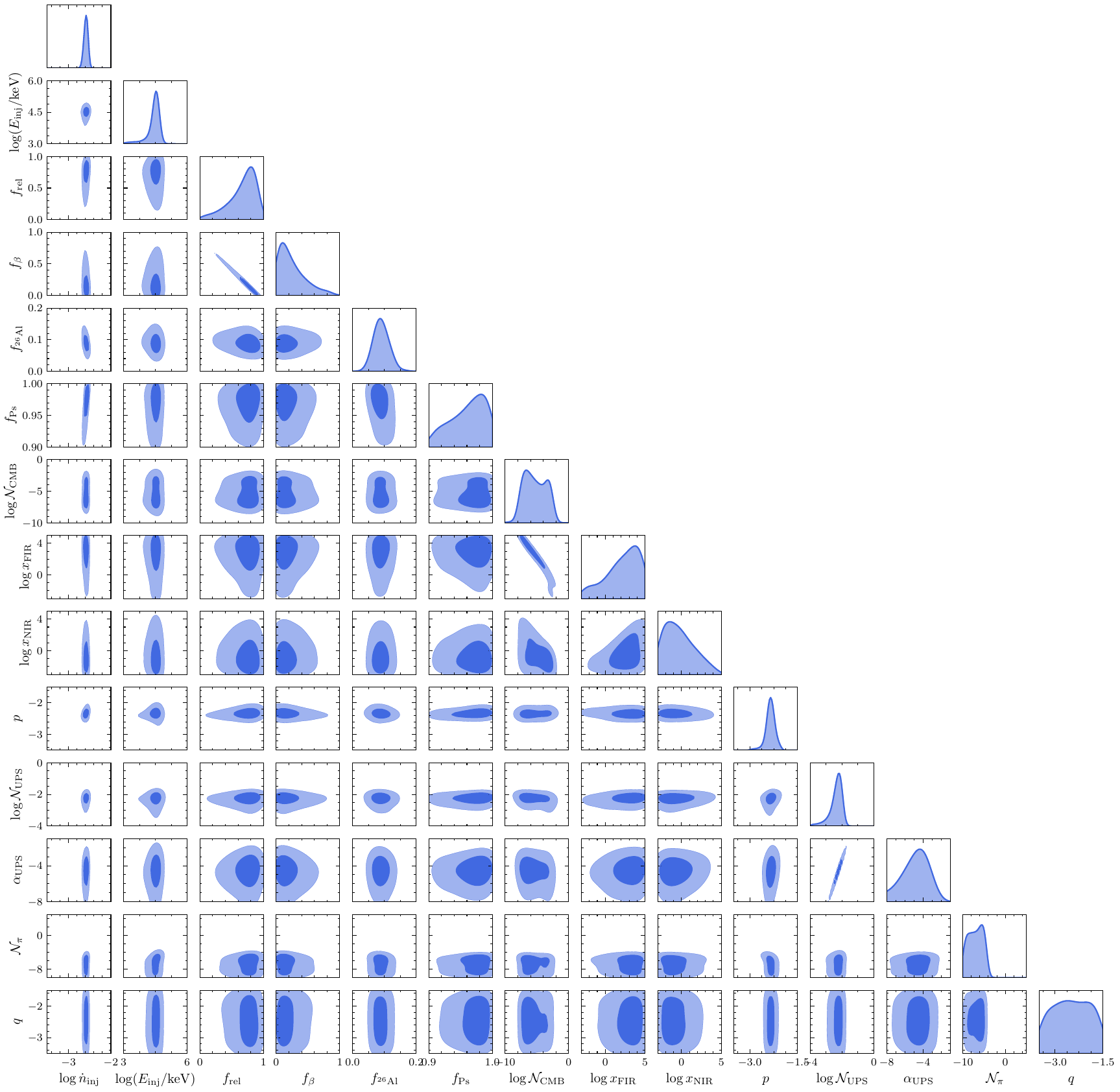}
    \caption{Full posteriors for (IV) ionized medium, $f_{\rm IC}=10^{-8}$, for the 18\degree~ROI.}
    \label{fig:fullposterior_ionized_8.0_18}
\end{figure*}

\begin{figure*}[!ht]
    \centering
    \includegraphics[width=1.0\linewidth]{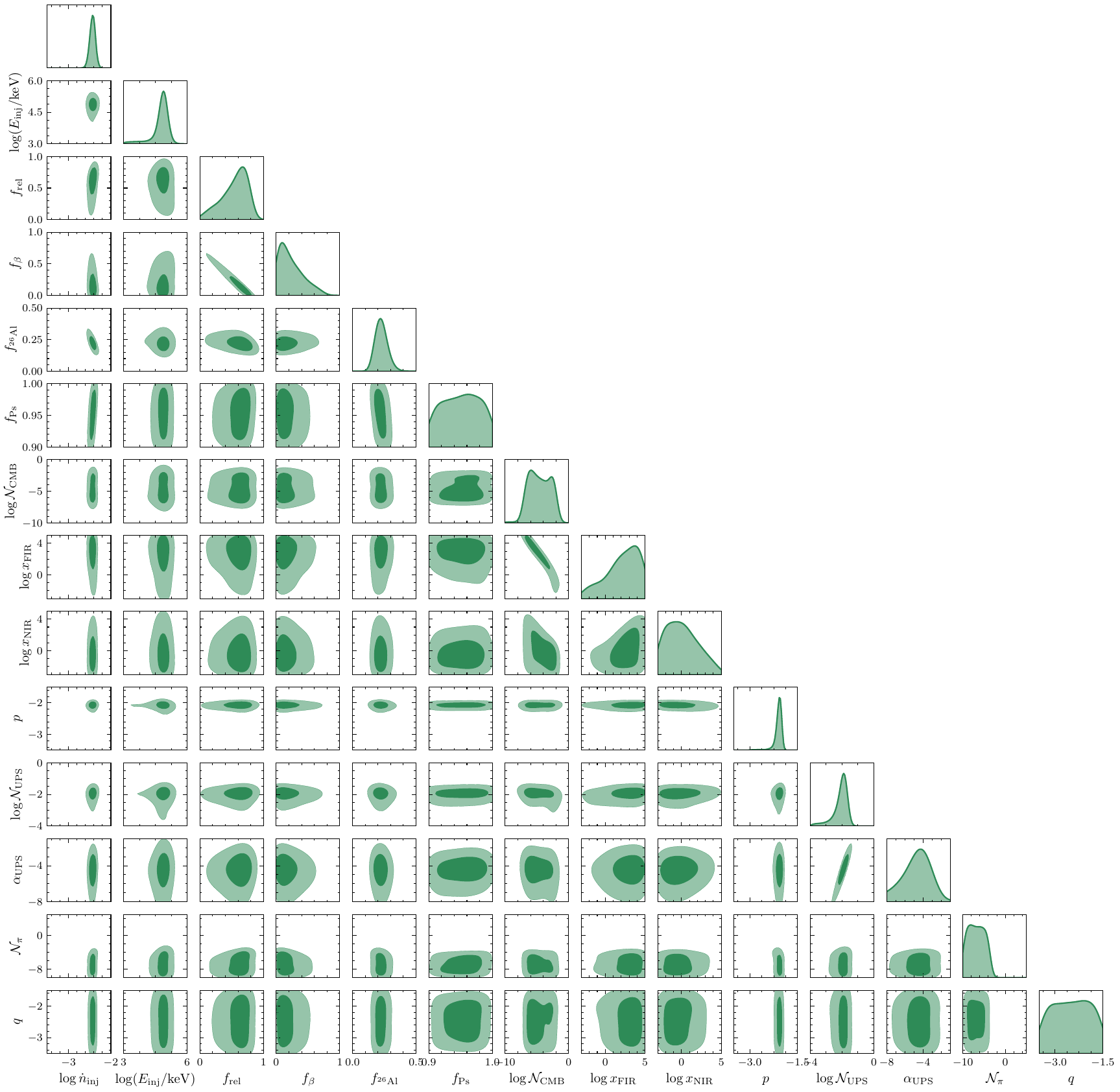}
    \caption{Full posteriors for (IV) ionized medium, $f_{\rm IC}=10^{-8}$, for the 95\degree~ROI.}
    \label{fig:fullposterior_ionized_8.0_95}
\end{figure*}

\begin{figure*}[!ht]
    \centering
    \includegraphics[width=1.0\linewidth]{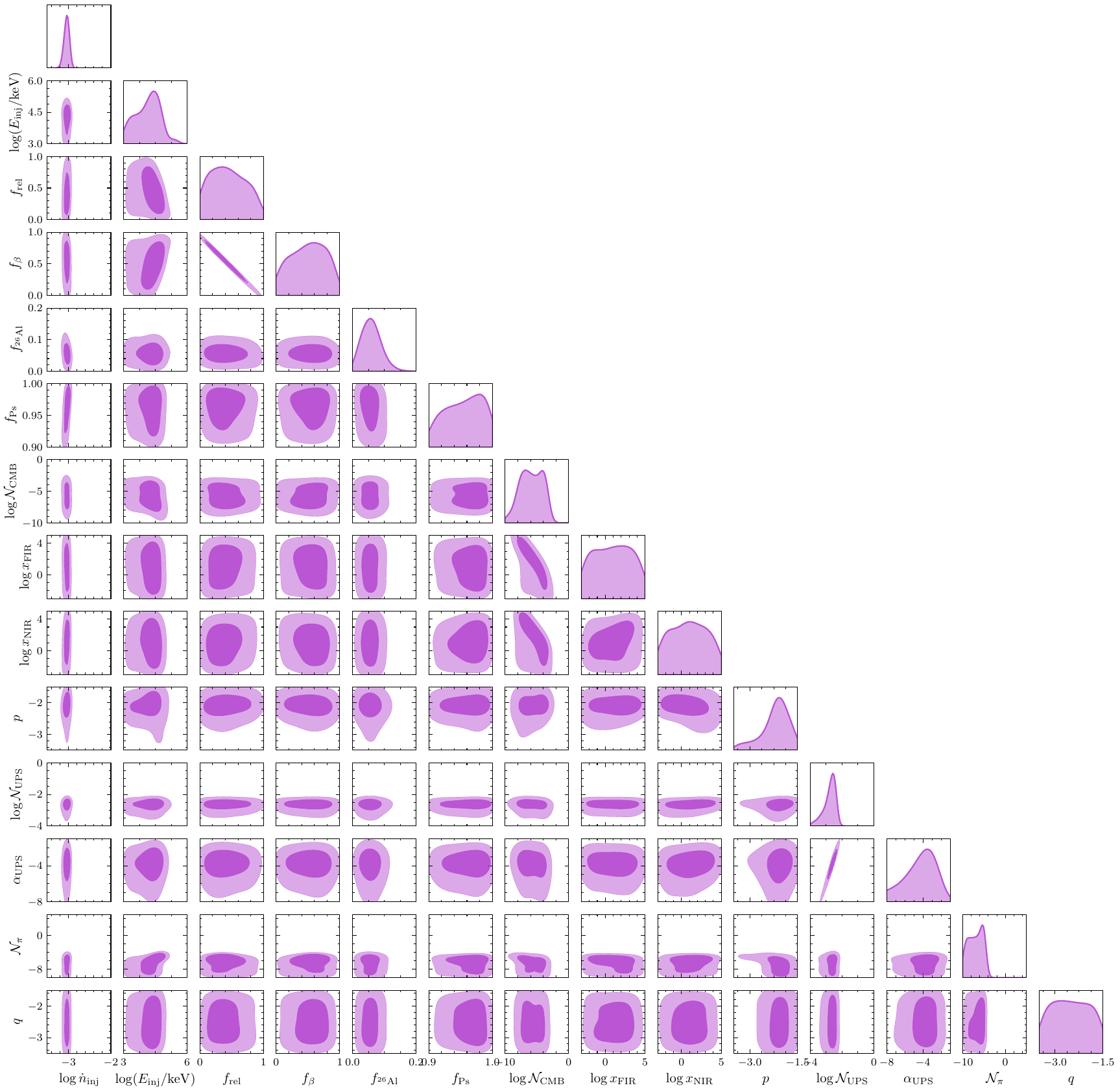}
    \caption{Full posteriors for (V) atomic medium, $f_{\rm IC}=10^{-6}$, for the 5\degree~ROI.}
    \label{fig:fullposterior_atomic_6.0_05}
\end{figure*}

\begin{figure*}[!ht]
    \centering
    \includegraphics[width=1.0\linewidth]{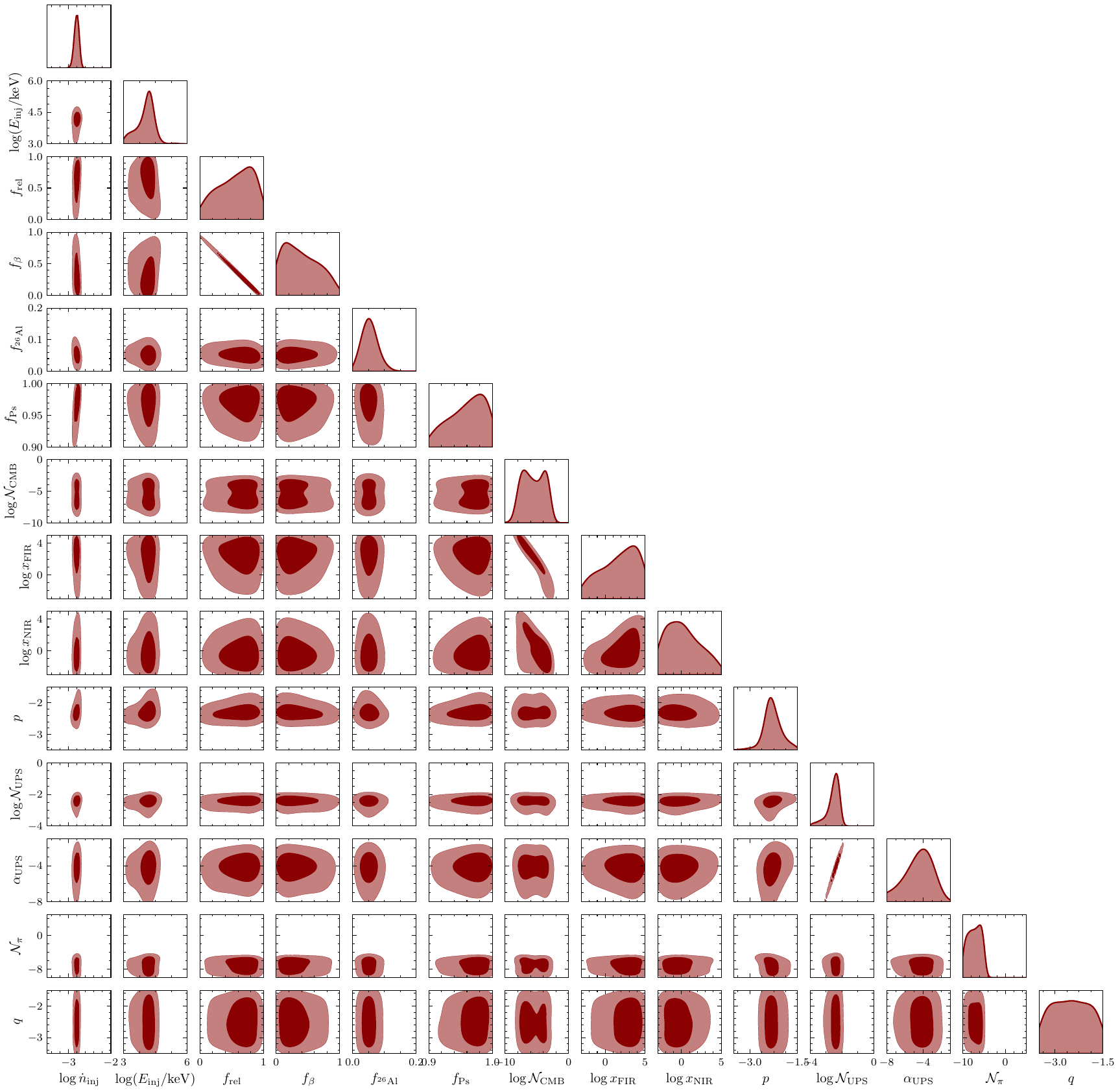}
    \caption{Full posteriors for (V) atomic medium, $f_{\rm IC}=10^{-6}$, for the 9\degree~ROI.}
    \label{fig:fullposterior_atomic_6.0_09}
\end{figure*}

\begin{figure*}[!ht]
    \centering
    \includegraphics[width=1.0\linewidth]{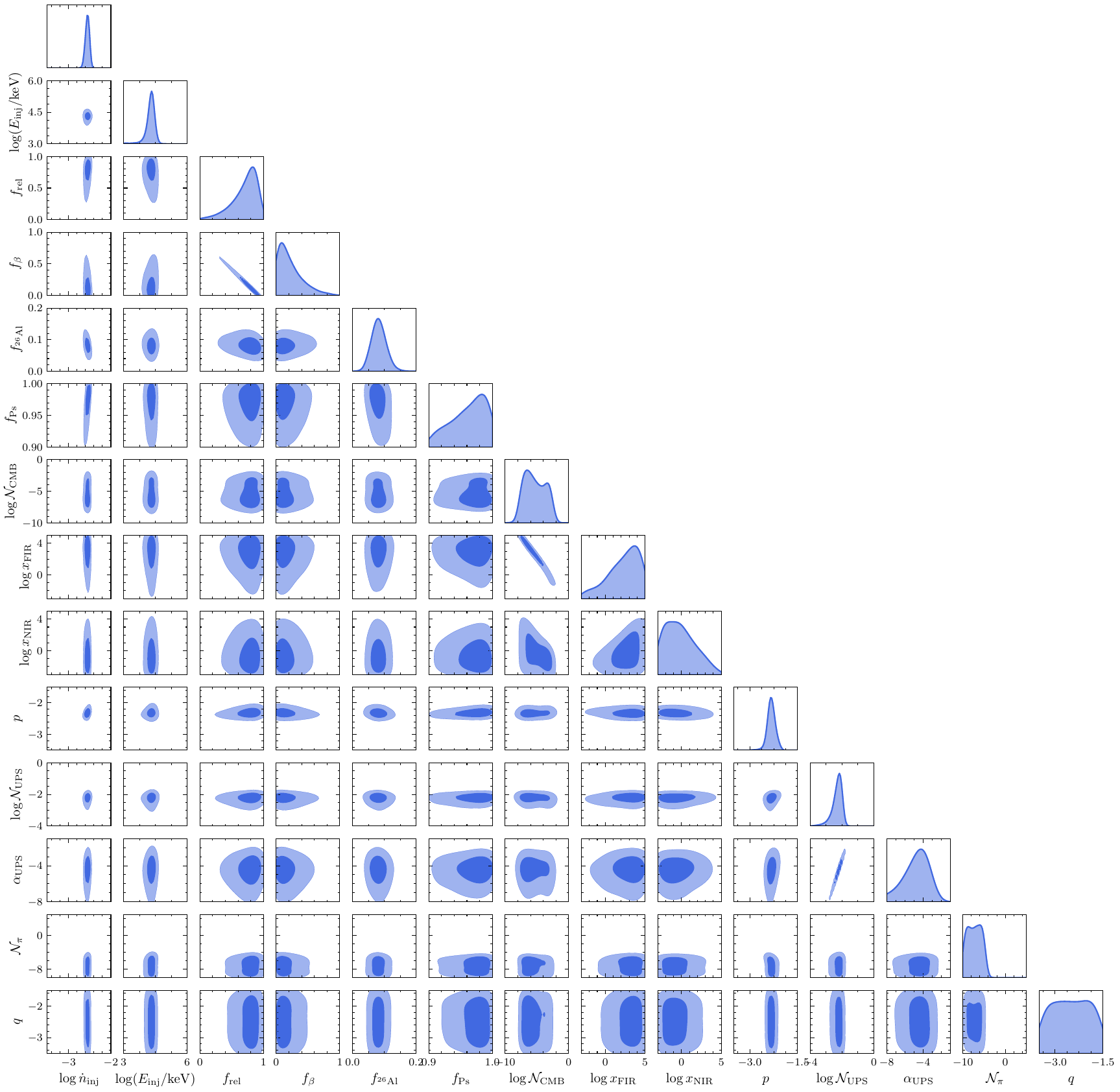}
    \caption{Full posteriors for (V) atomic medium, $f_{\rm IC}=10^{-6}$, for the 18\degree~ROI.}
    \label{fig:fullposterior_atomic_6.0_18}
\end{figure*}

\begin{figure*}[!ht]
    \centering
    \includegraphics[width=1.0\linewidth]{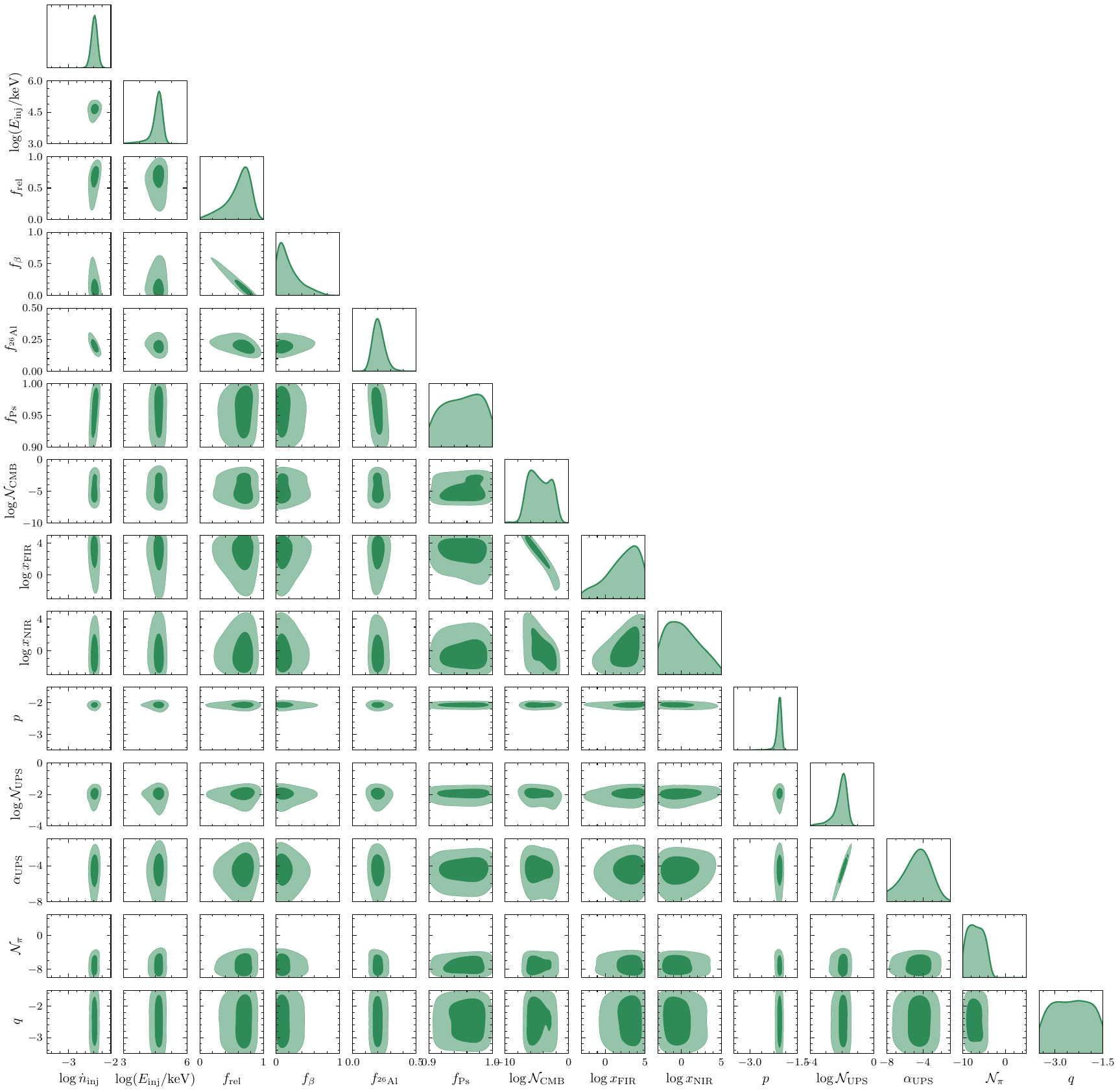}
    \caption{Full posteriors for (V) atomic medium, $f_{\rm IC}=10^{-6}$, for the 95\degree~ROI.}
    \label{fig:fullposterior_atomic_6.0_95}
\end{figure*}

In this section, we present the posteriors on the model parameters, obtained from MCMC analysis of the highest-relative-probability scaling for each ROI, as described in End Matter of the main paper. These are reported for the fiducial ISM as well as alternative choices (see above). The posteriors for each of the cases are summarized in Table \ref{tab:posteriors}, the corresponding spectral fits in Figures \ref{fig:spectral_fits_ionized_5.0}-\ref{fig:spectral_fits_atomic_6.0} (analogous to Figure \ref{fig:bestfit-ptsrc-09} of the main paper), and the corner plots for all variables and all combinations of ROI and ISM model appear in Figures \ref{fig:fullposterior_ionized_5.0_05}-\ref{fig:fullposterior_atomic_6.0_95}.

\renewcommand{\arraystretch}{1.6}
\begin{sidewaystable*}[h]
{
\footnotesize{
    \centering
    \vspace*{9cm}
    \begin{tabular}{ c c c c c c c c c c c c c c c } \hline \hline
 & $\log{\dot{n}_{\rm inj}}$ & $E_{\rm inj}$ [MeV] & $f_{\rm rel}$ & $f_{\beta}$ & $f_{\rm ^{26}Al}$ & $f_{\rm Ps}$ & $\log\mathcal{N}_{\rm CMB}$ & $\log x_{\rm FIR}$ & $\log x_{\rm NIR}$ & $p$ & $\log\mathcal{N}_{\rm UPS}$ & $\alpha_{\rm UPS}$ & $\log\mathcal{N}_{\pi}$ & $q$\\

 \multicolumn{15}{l}{\textbf{ROI = 5\degree circle}} \\ \hline
I & $-3.06^{+0.09}_{-0.10}$ & $39.2^{+35.4}_{-37.5}$ & $0.52^{+0.37}_{-0.43}$ & $0.42^{+0.43}_{-0.36}$ & $0.059^{+0.055}_{-0.044}$ & $0.96^{+0.04}_{-0.05}$ & $-6.54^{+2.96}_{-2.90}$ & $0.84^{+3.68}_{-3.46}$ & $0.81^{+3.68}_{-3.41}$ & $-2.16^{+0.55}_{-1.10}$ & $-2.8^{+0.4}_{-0.95}$ & $-3.9^{+2.7}_{-5}$ & $-5.84^{+1.03}_{-3.76}$ & $-2.51^{+0.90}_{-0.90}$\\ 
II$^*$ & $-3.06^{+0.08}_{-0.10}$ & $27.5^{+57.1}_{-26.1}$ & $0.49^{+0.40}_{-0.44}$ & $0.44^{+0.44}_{-0.40}$ & $0.059^{+0.055}_{-0.045}$ & $0.96^{+0.04}_{-0.05}$ & $-6.12^{+2.61}_{-2.89}$ & $0.91^{+3.63}_{-3.54}$ & $0.84^{+3.64}_{-3.46}$ & $-2.13^{+0.52}_{-1.05}$ & $-2.7^{+0.32}_{-0.92}$ & $-4.1^{+1.9}_{-5}$ & $-6.53^{+1.70}_{-3.13}$ & $-2.49^{+0.89}_{-0.90}$\\
III & $-3.06^{+0.09}_{-0.10}$ & $25.4^{+58.3}_{-23.9}$ & $0.49^{+0.40}_{-0.44}$ & $0.45^{+0.44}_{-0.40}$ & $0.06^{+0.055}_{-0.045}$ & $0.96^{+0.04}_{-0.05}$ & $-5.99^{+2.49}_{-2.97}$ & $1.09^{+3.46}_{-3.67}$ & $0.65^{+3.76}_{-3.30}$ & $-2.13^{+0.52}_{-1.00}$ & $-2.7^{+0.33}_{-0.87}$ & $-4.2^{+2}_{-4.8}$ & $-6.63^{+1.80}_{-3.06}$ & $-2.50^{+0.90}_{-0.90}$\\
IV & $-3.06^{+0.08}_{-0.10}$ & $25.9^{+55.1}_{-24.4}$ & $0.49^{+0.39}_{-0.44}$ & $0.44^{+0.44}_{-0.39}$ & $0.06^{+0.053}_{-0.044}$ & $0.96^{+0.04}_{-0.05}$ & $-6.09^{+2.58}_{-2.89}$ & $0.99^{+3.55}_{-3.56}$ & $0.75^{+3.77}_{-3.35}$ & $-2.13^{+0.52}_{-1.03}$ & $-2.7^{+0.33}_{-0.94}$ & $-4.1^{+1.9}_{-5.2}$ & $-6.61^{+1.78}_{-3.07}$ & $-2.51^{+0.91}_{-0.89}$\\
V & $-3.04^{+0.09}_{-0.11}$ & $16.4^{+55.9}_{-15}$ & $0.42^{+0.46}_{-0.38}$ & $0.52^{+0.38}_{-0.45}$ & $0.057^{+0.054}_{-0.043}$ & $0.96^{+0.04}_{-0.05}$ & $-5.85^{+2.34}_{-2.70}$ & $0.97^{+3.57}_{-3.61}$ & $0.89^{+3.61}_{-3.49}$ & $-2.12^{+0.50}_{-1.03}$ & $-2.7^{+0.33}_{-0.88}$ & $-4.1^{+1.9}_{-4.9}$ & $-6.91^{+2.05}_{-2.78}$ & $-2.51^{+0.91}_{-0.89}$\\ \hline

 \multicolumn{15}{l}{\textbf{ROI = 9\degree circle}} \\ \hline
I & $-2.83^{+0.07}_{-0.09}$ & $23.2^{+33.4}_{-21.6}$ & $0.58^{+0.34}_{-0.48}$ & $0.37^{+0.48}_{-0.33}$ & $0.055^{+0.046}_{-0.038}$ & $0.97^{+0.03}_{-0.06}$ & $-5.93^{+2.71}_{-2.66}$ & $1.90^{+2.81}_{-4.33}$ & $-0.15^{+4.03}_{-2.56}$ & $-2.33^{+0.62}_{-0.68}$ & $-2.5^{+0.32}_{-0.84}$ & $-4.4^{+1.9}_{-4.6}$ & $-6.77^{+1.92}_{-2.90}$ & $-2.49^{+0.89}_{-0.90}$\\
II$^*$ & $-2.83^{+0.08}_{-0.09}$ & $22.2^{+31.6}_{-20.6}$ & $0.57^{+0.33}_{-0.49}$ & $0.37^{+0.49}_{-0.33}$ & $0.057^{+0.046}_{-0.039}$ & $0.96^{+0.03}_{-0.05}$ & $-5.69^{+2.49}_{-2.64}$ & $1.88^{+2.84}_{-4.35}$ & $-0.27^{+4.05}_{-2.45}$ & $-2.33^{+0.58}_{-0.58}$ & $-2.5^{+0.32}_{-0.91}$ & $-4.4^{+1.8}_{-4.9}$ & $-6.89^{+2.03}_{-2.79}$ & $-2.51^{+0.91}_{-0.89}$\\ 
III & $-2.83^{+0.08}_{-0.09}$ & $22.2^{+31.6}_{-20.6}$ & $0.57^{+0.33}_{-0.49}$ & $0.37^{+0.49}_{-0.33}$ & $0.057^{+0.046}_{-0.039}$ & $0.96^{+0.03}_{-0.05}$ & $-5.69^{+2.49}_{-2.64}$ & $1.88^{+2.84}_{-4.35}$ & $-0.27^{+4.05}_{-2.45}$ & $-2.33^{+0.58}_{-0.58}$ & $-2.5^{+0.32}_{-0.91}$ & $-4.4^{+1.8}_{-4.9}$ & $-6.89^{+2.03}_{-2.79}$ & $2.51^{+0.89}_{-0.91}$\\  
IV & $-2.83^{+0.08}_{-0.09}$ & $21.5^{+32.3}_{-19.9}$ & $0.57^{+0.33}_{-0.50}$ & $0.37^{+0.49}_{-0.33}$ & $0.056^{+0.045}_{-0.039}$ & $0.96^{+0.03}_{-0.05}$ & $-5.79^{+2.57}_{-2.51}$ & $2.05^{+2.68}_{-4.41}$ & $-0.20^{+4.03}_{-2.52}$ & $-2.33^{+0.60}_{-0.58}$ & $-2.5^{+0.32}_{-0.9}$ & $-4.4^{+1.8}_{-4.9}$ & $-7.06^{+2.19}_{-2.65}$ & $-2.49^{+0.89}_{-0.91}$\\
V & $-2.81^{+0.08}_{-0.10}$ & $13.6^{+21}_{-11.9}$ & $0.59^{+0.33}_{-0.51}$ & $0.35^{+0.51}_{-0.32}$ & $0.053^{+0.044}_{-0.037}$ & $0.97^{+0.03}_{-0.05}$ & $-5.62^{+2.40}_{-2.32}$ & $2.14^{+2.60}_{-4.51}$ & $-0.02^{+3.98}_{-2.66}$ & $-2.31^{+0.53}_{-0.38}$ & $-2.5^{+0.32}_{-0.97}$ & $-4.5^{+1.8}_{-5.3}$ & $-7.29^{+2.14}_{-2.43}$ & $-2.52^{+0.91}_{-0.89}$\\ \hline

 \multicolumn{15}{l}{\textbf{ROI = 18\degree circle}} \\ \hline
I& $-2.59^{+0.07}_{-0.08}$ & $33^{+21.1}_{-28}$ & $0.72^{+0.18}_{-0.49}$ & $0.19^{+0.48}_{-0.18}$ & $0.089^{+0.041}_{-0.036}$ & $0.97^{+0.03}_{-0.06}$ & $-5.44^{+2.60}_{-2.16}$ & $2.59^{+2.19}_{-4.65}$ & $-0.48^{+3.76}_{-2.27}$ & $-2.36^{+0.23}_{-0.30}$ & $-2.3^{+0.32}_{-0.96}$ & $-4.8^{+1.8}_{-5.1}$ & $-6.84^{+2.27}_{-2.84}$ & $-2.50^{+0.90}_{-0.90}$\\ 
II$^*$ & $-2.59^{+0.07}_{-0.08}$ & $32.2^{+19}_{-27.5}$ & $0.72^{+0.18}_{-0.51}$ & $0.19^{+0.50}_{-0.17}$ & $0.089^{+0.042}_{-0.036}$ & $0.97^{+0.03}_{-0.06}$ & $-5.46^{+2.63}_{-2.12}$ & $2.62^{+2.18}_{-4.62}$ & $-0.48^{+3.76}_{-2.27}$ & $-2.36^{+0.22}_{-0.28}$ & $-2.3^{+0.32}_{-0.92}$ & $-4.8^{+1.8}_{-5}$ & $-6.98^{+2.39}_{-2.73}$ & $-2.52^{+0.92}_{-0.88}$\\
III & $-2.59^{+0.07}_{-0.08}$ & $32.3^{+18.7}_{-27.4}$ & $0.72^{+0.18}_{-0.50}$ & $0.19^{+0.49}_{-0.18}$ & $0.089^{+0.042}_{-0.036}$ & $0.97^{+0.03}_{-0.06}$ & $-5.54^{+2.67}_{-2.03}$ & $2.74^{+2.07}_{-4.56}$ & $-0.50^{+3.76}_{-2.26}$ & $-2.35^{+0.22}_{-0.26}$ & $-2.3^{+0.31}_{-0.82}$ & $-4.7^{+1.7}_{-4.5}$ & $-6.96^{+2.34}_{-2.73}$ & $-2.50^{+0.90}_{-0.90}$\\
IV & $-2.59^{+0.07}_{-0.08}$ & $32.1^{+17.7}_{-26.9}$ & $0.73^{+0.17}_{-0.50}$ & $0.18^{+0.49}_{-0.17}$ & $0.089^{+0.041}_{-0.036}$ & $0.97^{+0.03}_{-0.06}$ & $-5.38^{+2.56}_{-2.16}$ & $2.60^{+2.18}_{-4.70}$ & $-0.58^{+3.76}_{-2.17}$ & $-2.35^{+0.21}_{-0.25}$ & $-2.3^{+0.32}_{-0.93}$ & $-4.8^{+1.8}_{-5}$ & $-6.95^{+2.30}_{-2.76}$ & $-2.49^{+0.89}_{-0.90}$\\
V & $-2.55^{+0.07}_{-0.08}$ & $20^{+12.1}_{-12.4}$ & $0.76^{+0.16}_{-0.43}$ & $0.16^{+0.43}_{-0.15}$ & $0.082^{+0.039}_{-0.034}$ & $0.97^{+0.03}_{-0.06}$ & $-5.41^{+2.56}_{-2.10}$ & $2.69^{+2.11}_{-4.43}$ & $-0.43^{+3.73}_{-2.32}$ & $-2.33^{+0.21}_{-0.19}$ & $-2.2^{+0.3}_{-0.69}$ & $-4.7^{+1.6}_{-3.8}$ & $-7.29^{+2.38}_{-2.45}$ & $-2.51^{+0.90}_{-0.89}$\\ \hline

 \multicolumn{15}{l}{\textbf{ROI = 95\degree square}} \\ \hline
I & $-2.42^{+0.10}_{-0.12}$ & $58.5^{+14.3}_{-54.6}$ & $0.65^{+0.16}_{-0.47}$ & $0.13^{+0.44}_{-0.12}$ & $0.22^{+0.09}_{-0.067}$ & $0.95^{+0.04}_{-0.05}$ & $-4.81^{+2.57}_{-1.98}$ & $2.79^{+2.00}_{-3.84}$ & $0.05^{+3.76}_{-2.76}$ & $-2.06^{+0.06}_{-0.20}$ & $-2.3^{+0.5}_{-1.4}$ & $-4.8^{+4}_{-6.1}$ & $-6.99^{+2.75}_{-2.72}$ & $-2.50^{+0.90}_{-0.89}$\\ 
II$^*$ & $-2.43^{+0.10}_{-0.11}$ & $66.4^{+32.7}_{-60.5}$ & $0.62^{+0.18}_{-0.44}$ & $0.15^{+0.43}_{-0.14}$ & $0.22^{+0.087}_{-0.066}$ & $0.95^{+0.04}_{-0.05}$ & $-4.72^{+2.49}_{-2.11}$ & $2.63^{+2.16}_{-4.59}$ & $-0.01^{+3.76}_{-2.67}$ & $-2.07^{+0.06}_{-0.23}$ & $-2^{+0.34}_{-1.1}$ & $-4.9^{+1.9}_{-5.5}$ & $-6.91^{+2.88}_{-2.79}$ & $-2.48^{+0.88}_{-0.91}$\\
III & $-2.43^{+0.10}_{-0.11}$ & $71.6^{+51.8}_{-64.6}$ & $0.59^{+0.20}_{-0.43}$ & $0.18^{+0.41}_{-0.16}$ & $0.22^{+0.09}_{-0.066}$ & $0.95^{+0.04}_{-0.05}$ & $-4.71^{+2.47}_{-2.09}$ & $2.66^{+2.12}_{-4.37}$ & $-0.11^{+3.77}_{-2.59}$ & $-2.07^{+0.07}_{-0.21}$ & $-2^{+0.31}_{-0.97}$ & $-4.8^{+1.7}_{-5.2}$ & $-6.82^{+2.87}_{-2.87}$ & $-2.47^{+0.88}_{-0.92}$\\ 
IV & $-2.43^{+0.10}_{-0.11}$ & $71.8^{+55.4}_{-65.5}$ & $0.59^{+0.19}_{-0.43}$ & $0.18^{+0.41}_{-0.16}$ & $0.22^{+0.087}_{-0.066}$ & $0.95^{+0.04}_{-0.05}$ & $-4.66^{+2.42}_{-2.15}$ & $2.59^{+2.19}_{-4.45}$ & $-0.14^{+3.74}_{-2.56}$ & $-2.08^{+0.07}_{-0.25}$ & $-2^{+0.31}_{-1.1}$ & $-4.7^{+1.8}_{-5.5}$ & $-6.81^{+2.85}_{-2.87}$ & $-2.48^{+0.89}_{-0.91}$\\
V & $-2.39^{+0.11}_{-0.12}$ & $43.6^{+27.1}_{-36.1}$ & $0.65^{+0.16}_{-0.43}$ & $0.14^{+0.40}_{-0.13}$ & $0.2^{+0.086}_{-0.06}$ & $0.95^{+0.04}_{-0.05}$ & $-4.71^{+2.50}_{-2.07}$ & $2.64^{+2.15}_{-4.44}$ & $-0.11^{+3.89}_{-2.59}$ & $-2.07^{+0.06}_{-0.20}$ & $-2^{+0.31}_{-1}$ & $-4.7^{+1.8}_{-5.3}$ & $-6.99^{+2.84}_{-2.70}$ & $-2.49^{+0.90}_{-0.91}$\\ \hline
\multicolumn{15}{l}{$^*$: Benchmark ISM model}\\
\multicolumn{15}{p{1.0\textwidth}}{
$^\#$: The quantities $\dot{n}_{\rm inj}$, $\mathcal{N}_{\rm CMB}$, $\mathcal{N}_{\rm UPS}$, and $\mathcal{N}_{\pi}$ are normalized to the respective squared effective distances $4\pi D_{\rm inj}^2$, $4\pi D_{\rm CMB}^2$, $4\pi D_{\rm NIR}^2$ and $4\pi D_{\pi}^2$, and in units of $\rm{cm^{-2}~ s^{-1}}$ as described in Table \ref{tab:priors} of the main paper.}\\
    \end{tabular}
    \caption{Posteriors on model parameters in the various choices of ISM composition (I-V) as described earlier. The posteriors reported for each case correspond to the scaling model with the highest relative probability (see Table \ref{tab:bayes}). The median value of the posterior on each parameter is reported along with its difference from the 5th and 95th percentiles.}
    \label{tab:posteriors}}}
\end{sidewaystable*}

\renewcommand{\floatpagefraction}{.8}
\renewcommand{\topfraction}{.85}
\renewcommand{\bottomfraction}{.7}
\renewcommand{\textfraction}{.15}
\setcounter{totalnumber}{5}
\setcounter{topnumber}{5}
\setcounter{bottomnumber}{5}









\clearpage

\end{document}